\documentclass[a4paper,11pt]{article}
\usepackage{jheppub} 

\usepackage[caption=false]{subfig}
\usepackage{color}

\definecolor{teal}{rgb}{0.0, 0.5, 0.5}

\usepackage{setspace}

\usepackage[normalem]{ulem}

\newcommand{\sven}{\textsc{Sven}}
\newcommand{\mathematica}{\textsc{Mathematica}}

\usepackage{placeins} 

\DeclareRobustCommand{\Sec}[1]{Sec.~\ref{sec:#1}}

\DeclareRobustCommand{\App}[1]{App.~\ref{app:#1}}

\DeclareRobustCommand{\Tab}[1]{Table~\ref{tab:#1}}
\DeclareRobustCommand{\Tabs}[2]{Tables~\ref{tab:#1} and \ref{tab:#2}}
\DeclareRobustCommand{\Fig}[1]{Fig.~\ref{fig:#1}}
\DeclareRobustCommand{\Figs}[2]{Figs.~\ref{fig:#1} and \ref{fig:#2}}

\DeclareRobustCommand{\Eq}[1]{Eq.~(\ref{eq:#1})}

\DeclareRobustCommand{\Reference}[1]{Ref.~\cite{#1}}
\DeclareRobustCommand{\References}[1]{Refs.~\cite{#1}}

\title{\boldmath Descending into the Modular Bootstrap}

\author[a]{Nathan Benjamin,}
\author[b]{A. Liam Fitzpatrick,}
\author[b]{Wei Li,}
\author[c,d,e,f]{and Jesse Thaler}

\preprint{MIT-CTP/6023}

\affiliation[a]{Department of Physics and Astronomy, University of Southern California, Los Angeles, CA 90089, USA}
\affiliation[b]{Department of Physics, Boston University, Boston, MA 02215, USA}
\affiliation[c]{Center for Theoretical Physics -- a Leinweber Institute, Massachusetts Institute of Technology,
Cambridge, MA 02139, USA}
\affiliation[d]{Institut des Hautes \'Etudes Scientifiques, 91440 Bures-sur-Yvette, France}
\affiliation[e]{Institut de Physique Th\'eorique, CEA Paris-Saclay, 91191 Gif-sur-Yvette, France}
\affiliation[f]{The NSF Institute for Artificial Intelligence and Fundamental Interactions}

\emailAdd{nathanbe@usc.edu}
\emailAdd{fitzpatr@bu.edu}
\emailAdd{weili17@bu.edu}
\emailAdd{jthaler@mit.edu}

\abstract{
In this paper, we attempt to explore the landscape of two-dimensional conformal field theories (2d CFTs) by efficiently searching for numerical solutions to the modular bootstrap equation using machine-learning-style optimization.
The torus partition function of a 2d CFT is fixed by the spectrum of its primary operators and its chiral algebra, which we take to be the Virasoro algebra with $c>1$.
We translate the requirement that this partition function is modular invariant into a loss function, which we then minimize to identify possible primary spectra.
Our approach involves two technical innovations that facilitate finding reliable candidate CFTs.
The first is a strategy to estimate the uncertainty associated with truncating the spectrum to the lowest dimension operators.
The second is the use of a new singular-value-based optimizer (\sven) that is more effective than gradient descent at navigating the hierarchical structure of the loss landscape.
We numerically construct candidate truncated CFT partition functions with central charges between 1 and $\frac{8}{7}$, a range devoid of known examples, and argue that these candidates likely come from a continuous space of modular bootstrap solutions.
We also provide evidence for a more stringent constraint on the spectral gap near $c = 1$ than the existing bound of $\Delta_{\rm gap} \le \frac{c}{6} + \frac{1}{3}$.}

\begin{document}

\maketitle
\flushbottom

\section{Introduction}

Are consistent quantum field theories (QFTs) rare or abundant?
If they are rare, then it makes sense to exhaustively catalog them.
If they are abundant, though, any individual QFT tells you little about the space of consistent QFTs, and it makes more sense to establish broad properties of and constraints on such systems.
At least within perturbation theory, Lagrangians provide a systematic and practical way to explore such questions.
More generally, though, we are still far from any kind of explicit catalog of all QFTs at strong coupling.

The case of two-dimensional conformal field theories (2d CFTs) offers at least a clear mathematical framework for constructing such a catalog.
CFTs are defined by a discrete countable set of data, namely the spectrum of conformal weights and operator product expansion (OPE) coefficients of primary operators, together with a chiral algebra.%
\footnote{More generally, non-compact CFTs like Liouville CFT can have continuous spectra and lack an $\text{SL}(2,\mathbb R)$ invariant vacuum state. We will only consider compact CFTs in this work, by insisting there is a vacuum in the spectrum.}
Moreover, there is a complete set of axioms that 2d CFTs have to satisfy, the most important of which are modular invariance of the torus partition function, crossing symmetry of the sphere four-point correlators, and modular covariance of the torus one-point functions.
Thus, the space of 2d CFTs can concretely be considered the space of sets of CFT data that satisfy the equations imposed by these constraints.
Finding such solutions is still an extremely challenging problem, but it is at least possible to explore the space of such solutions and their properties numerically.
This is in fact what the conformal bootstrap and modular bootstrap equations do, by finding general constraints imposed by these conditions.

In this paper, we take a complementary approach, sometimes called the ``primal'' approach, to these constraints, where we attempt to explore the space of possible 2d CFTs by explicitly constructing solutions to a subset of the full set of constraints.
We only consider the constraint of modular invariance of the torus partition function, and therefore it is more accurate to say that we are exploring the space of 2d CFT partition functions rather than the space of 2d CFTs.
 Because modular invariance imposes an effectively infinite number of constraints, we will at best be able to construct approximate solutions.
 The exact definition of ``approximate'' that we use will be one of the main features of this paper, and we present evidence that key features of the space of approximate solutions we find are robust to improving the approximation.
A grand (and perhaps quixotic) long-term goal of our approach would be to ultimately find approximate solutions to all the 2d CFT constraints and thereby {\it construct} new, fully-fledged 2d CFTs.

At $c = 1$, there is an exhaustive catalog of unitary 2d CFTs,%
\footnote{It is still unproven that the known $c=1$ CFTs are exhaustive, though they are widely believed to be.}
including a moduli space of compactified boson theories, their orbifolds, and a discrete set of exceptional cases~\cite{Ginsparg:1988ui, Ginsparg:1987eb}.
For $c > 1$, no exhaustive catalog is known, though a number of properties and constraints have already been established, including the celebrated Cardy formula on the asymptotic density of states~\cite{Cardy:1986ie}, various operator bounds derived through the modular bootstrap~\cite{Cardy:1991kr,Hellerman:2009bu,Friedan:2013cba,Collier:2016cls, Hartman:2019pcd}, and holographic relations at very large $c$~\cite{Strominger:1997eq,Hartman:2014oaa,Fitzpatrick:2014vua}.
There are several infinite classes of constructions of 2d CFTs at $c>1$.
The Wess-Zumino-Witten (WZW) coset models (which include the minimal series CFTs as special cases) are rational CFTs, meaning that they have a finite number of (chiral) primary operators, and have simple, explicit formulas for their operator spectrum.
Additionally, there are continuous families from Narain moduli space and Calabi-Yau moduli space, based on a geometric ``target space'' interpretation.
Nevertheless, despite this large set of known 2d CFTs, they are all ``non-generic'' in the sense that they all are either rational (have a finite number of primary operators under the theory's chiral algebra) or are marginal deformations of rational theories.
Indeed, although the standard lore is that ``most'' 2d CFTs should be irrational,%
\footnote{As far as we know, this lore is not based on any rigorous result but rather reflects the idea that solvable theories are special and chaotic theories are more generic; the fact that all known 2d CFTs are rational theories (or marginal deformations thereof) is likely a lamppost effect. Moreover, the conformal algebra forces unitary CFTs to be rational at sufficiently small central charge (e.g.~$c<1$ for non-supersymmetric CFTs, $c<\frac{3}{2}$ for ${\cal N}=1$ superconformal CFTs, etc.), but not at larger central charge.}
with an infinite number of (Virasoro) primaries and no conserved currents beyond the stress tensor, there are no completely explicit examples.%
\footnote{There are some very interesting recent attempts. See for example  \References{Antunes:2022vtb, Antunes:2024mfb, Antunes:2025erb}, which looked at short renormalization group flows from coupled minimal models using perturbation theory, and \Reference{Antunes:2025huk}, which constructed a lattice model that conjecturally flows to an irrational CFT with $c= 2.10 \pm 0.03$. See also \Reference{Li:2025czz} for another possible strategy of discovering new 2d CFTs.  For a review of earlier work on irrational CFTs
based on the Virasoro master equation, see \Reference{Halpern:1995js}; known constructions of this form have $c \gtrsim 1.74$, outside the range explored here.
}
Without any confirmed examples of irrational compact 2d CFTs with only Virasoro symmetry,
though, we have little to guide our intuition about whether such systems should be abundant, rare, or non-existent.

\begin{figure}
    \centering
    \subfloat[][]{
        \includegraphics[width=0.45\linewidth]{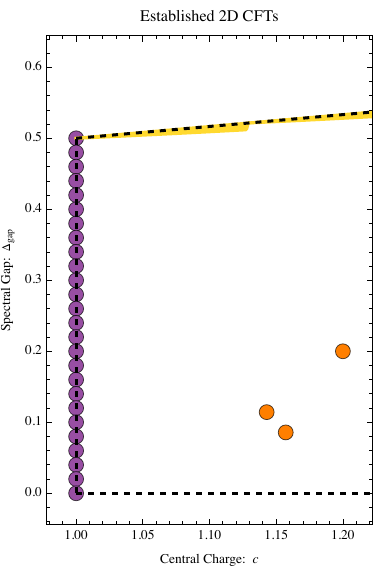}
    }$\qquad$
    \subfloat[][]{
        \includegraphics[width=0.45\linewidth]{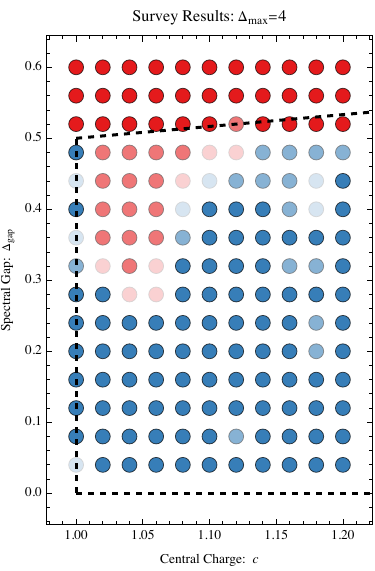}
        \label{fig:intro_yesno_examples}
    }
    \caption{Space of (a) established 2d CFTs compared to (b) the results of this study, in the $(c,\Delta_{\rm gap})$ plane of central charge versus spectral gap.
    The dashed black lines correspond to the unitarity bound $\Delta_{\rm gap} \ge 0$ and the known dual bound $\Delta_{\rm gap} \le \frac{c}{6}+\frac{1}{3}$~\cite{Hellerman:2009bu,Friedan:2013cba,Collier:2016cls}, while the yellow band indicates the refined exclusion from \Reference{Fitzpatrick:2023lvh}.
    For the established 2d CFTs, the purple dots represent compactified free boson theories at $c = 1$ which have spectral gaps that fill the entire $\Delta_{\rm gap}  \in (0,\frac{1}{2})$ range.
    The orange dots correspond, from left to right, to the $\mathbb{Z}_5$ parafermion CFT with $(c,\Delta_{\rm gap}) = (\frac{8}{7}, \frac{4}{35})$~\cite{Fateev:1985mm}, the $N=1$ minimal model at $m=5$ with $(c,\Delta_{\rm gap}) = (\frac{81}{70}, \frac{3}{35})$~\cite{Friedan:1984rv}, and the Ising $\otimes$ tricritical Ising theory with $(c,\Delta_{\rm gap}) =(\frac{6}{5},\frac{1}{5})$.
    For our optimization results, which are a preview of \Fig{survey_results}, the blue dots correspond to numerical solutions of the primal modular bootstrap with a truncation scale of $\Delta_{\rm max} = 4$, with dark/medium/light blue corresponding to high/medium/low confidence in the solution.
    The red dots indicate poor optimization results, with dark/medium/light red corresponding to strong/intermediate/weak exclusion of that parameter point.
    Despite multiple attempts, we were unable to find good solutions in the upper left corner of this plot with $c \in (1.00,1.06)$ and $\Delta_{\rm gap} \in (0.3,0.5)$, suggestive of a primal/dual gap.
    We suspect that this gap might even grow with larger values of $\Delta_{\rm max}$. 
    }
    \label{fig:summary_plots}
\end{figure}

In this paper, we focus on the central charge range of:
\begin{equation}
    \label{eq:central_charge_range}
    c \in \left(1, \frac{8}{7}\right),
\end{equation}
for which there are no confirmed 2d CFTs, exactly solvable  or otherwise.
(The end point at $c = \frac{8}{7} \approx 1.14$ corresponds to the $\mathbb{Z}_5$ parafermion CFT~\cite{Fateev:1985mm}, with the nearby $N=1$ minimal model with $m=5$ at $c = \frac{81}{70} \approx 1.16$~\cite{Friedan:1984rv}.)
Using a truncated version of the modular bootstrap equations, we use an optimization approach styled after machine learning (ML) to identify candidate spectra that are as ``CFT-like'' as known 2d CFTs.
To define what we mean by CFT-like, we construct a $\chi^2$-style loss function that involves a theoretical estimate of the truncation uncertainty on the Euclidean torus partition function.
To identify CFT-like solutions, we minimize this loss using a new singular value descent (\sven) algorithm~\cite{Bright-Thonney:2026tlr}, which is more efficient at navigating multi-scale loss landscapes than standard gradient descent. 
We present a handful of explicit candidate 2d CFT spectra, together with uncertainty estimates on their operator dimensions, which provide strong evidence that these candidates are part of a continuous space of (truncated) modular bootstrap solutions.
As previewed in \Fig{summary_plots}, we also find numerical evidence for a more stringent bound on the spectral gap $\Delta_{\rm gap}$ than the existing bound of $\Delta_{\rm gap} \le \frac{c}{6} + \frac{1}{3}$~\cite{Hellerman:2009bu,Friedan:2013cba,Collier:2016cls}\footnote{\Reference{Hellerman:2009bu} introduced the modular bootstrap and derived the bound $\Delta_{\rm gap} \le \frac{c}{6} + 0.473694978898$ for all $c>1$.
The analysis in \Reference{Friedan:2013cba} applied numeric semidefinite programming methods to the spinless modular bootstrap (i.e.~$\tau$ restricted to be pure imaginary) which improved the bound slightly, and \Reference{Collier:2016cls} applied similar methods to the full modular bootstrap to improve the bound to $\Delta_{\rm gap} \le \frac{c}{6}+\frac{1}{3}$ for all $c\in [1,4]$.} and its refinements~\cite{Fitzpatrick:2023lvh}.%
\footnote{A previous version of this paper extended the bounds from \Reference{Fitzpatrick:2023lvh} through a naive linear extrapolation.  A direct calculation revealed a notch around $c = 1.13$ that will be investigated in future work.}

There are three key limitations of the primal modular bootstrap that we want to emphasize.
First, as already mentioned, the modular bootstrap equations only impose necessary but not sufficient conditions on the properties of a 2d CFT.
For this reason, any solutions we find are at best ``candidate CFT partition functions'';  more detailed exploration of  full CFT consistency is left to future work.
Second, our primal approach is based on truncating the full modular bootstrap equations.
While we try to account for truncation uncertainties in our method as best we can, it is possible that some of the solutions we find would not persist if the truncation scale were increased.
Indeed, we find evidence that raising $\Delta_{\rm max}$ likely shrinks the space of candidate CFTs compared to what is shown in \Fig{intro_yesno_examples}.
Third, unlike the dual modular bootstrap, we cannot definitively rule out candidate 2d CFTs.
As the adage goes, the absence of evidence is not evidence of absence, so even if we cannot identify a solution using our primal approach, a solution might nevertheless exist.
With these limitations in mind, a key advantage of the primal approach is that we can constrain CFT primaries to have integer-valued degeneracy, which is more difficult to ensure in the dual approach (see however \References{Kaidi:2020ecu,Fitzpatrick:2023lvh}).
This feature will enable us to predict the dimensions of dozens of operators simultaneously, with a fascinating hierarchical and correlated uncertainty structure strongly indicative of a continuous space of modular invariant CFT partition functions with integer degeneracies.

There are two technical points worth emphasizing related to the ML aspects of our study.
\begin{itemize}
\item \textbf{Uncertainty Estimates.}
The solution to any ML-style problem is only as good as the loss function one is trying to optimize.
A key theoretical innovation in this paper is an estimate of the uncertainty due to truncating the operator spectrum, which allows us to construct a $\chi^2$-style loss function.
This estimate is based on deforming a known CFT to have a different effective central charge, inspired by the primary partition function defined in \Reference{Benjamin:2021ygh}.%
\footnote{\label{footnote:claude}We thank Claude Opus 4.1 for helpful discussions that led to this deformation strategy.}
Without this uncertainty estimate, we found that a standard mean-squared-error (MSE) loss yields highly unstable and unphysical solutions.
Normalizing the squared error by the truncation uncertainty estimate, we can more confidently claim that $\chi^2$-style loss values of order 1 correspond to CFT-like behavior.
\item \textbf{Hierarchical Loss Navigation.}
With sufficiently complex loss landscapes, one can often get trapped in suboptimal local minima.
Our loss function turns out to have an exponential hierarchy of scales, related to the exponential form of the Virasoro characters.
For this reason, standard gradient descent cannot find particularly deep loss minima.
By contrast, the \sven\ algorithm~\cite{Bright-Thonney:2026tlr} (partially inspired by the challenges of this study) is able to find good candidate CFTs, since it can simultaneously traverse multiple parameter directions determined by the singular values of a suitable Jacobian matrix.
(Roughly speaking, gradient descent only probes one parameter direction.)
Nevertheless, the default version of \sven\ also struggles to reach the order 1 loss target, so we introduced an additional optimization trick to more reliably identify our final candidate spectra.
\end{itemize}
Beyond the current study of 2d CFTs, we suspect that these two technical points will be relevant for confronting other theoretical physics problems amenable to ML-style optimization.
In general, there are only a few attempts to include uncertainties when training ML surrogates for theoretical calculations in the quantum field theory literature~(see \Reference{Bahl:2026qaf} for a recent discussion).
In the bootstrap literature, truncation-error estimates have been studied previously in \References{Pappadopulo:2012jk,Rychkov:2015lca,Qiao:2017xif,Mukhametzhanov:2018zja, Mukhametzhanov:2019pzy,Pal:2019zzr,Mukhametzhanov:2020swe, Niarchos:2023lot, Marchetto:2023xap,Niarchos:2025cdg}. 
The novelty of our approach lies in translating this kind of \emph{a priori} error estimate into a quasi-statistical uncertainty, without relying on a learned error model.

Before beginning our hunt for 2d CFT partition functions, it is worth mentioning other related research on numerical optimization for bootstrap problems.
The dual conformal bootstrap is well established (see \References{Poland:2018epd,Rychkov:2023wsd} for reviews), with many advances through the use of semi-definite programming.
There have been a handful of ML-style approaches to studying CFTs; the approach that is philosophically closest to ours is that of \Reference{Niarchos:2025cdg}, which optimizes a truncation-aware loss function to find primal solutions to the finite temperature bootstrap.
Other ML approaches include ones based on classification~\cite{Chen:2020dxg,Kuo:2021lvu}, reinforcement learning~\cite{Kantor:2021kbx,Kantor:2021jpz,Kantor:2022epi,Niarchos:2023lot}, generative modeling~\cite{Laio:2022ayq}, genetic algorithms~\cite{Huang:2025qkk}, and spectral bias studies~\cite{Ghosh:2026jbw,Ghosh:2026xnp}.
There are related gradient-based approaches to solving dual bootstrap problems~\cite{Reehorst:2021ykw,Afkhami-Jeddi:2021iuw}.
Attempts at constructively finding a solution to a truncated set of the bootstrap equations (the ``Gliozzi bootstrap'' \cite{Gliozzi:2013ysa,Gliozzi:2016cmg} and its generalizations) can be formulated as an optimization problem, but so far typically require that one is given a priori some model-specific information about the low-lying spectrum of the CFT of interest, in the form of specific operator dimensions or the sparsity of operators in a particular OPE \cite{Afkhami-Jeddi:2019zci,Poland:2023vpn,ArguelloCruz:2025zuq,Kantor:2021kbx,Hu:2025yrs,Esterlis:2016psv,Laio:2022ayq}.
Beyond CFTs, ML-style strategies have also been applied to the S-matrix bootstrap~\cite{Dersy:2023job,Mizera:2023bsw,Niarchos:2024onf,Bhat:2024agd,Gumus:2024lmj,Gumus:2026mhb}.

The rest of this paper is organized as follows.
In \Sec{review}, we review the standard modular bootstrap program, especially the constraints imposed on the Euclidean torus partition function of 2d CFTs from modular invariance.
In \Sec{ml}, we show how to translate the modular bootstrap into the language of ML, (re)introduce the \sven\ optimization algorithm, and discuss how to estimate uncertainties from derivatives of the loss.
In \Sec{truncation}, we present a novel strategy to estimate the impact of truncating the spectrum at a finite $\Delta_{\rm max}$, based on deforming a known CFT.
We validate our analysis procedure using $N=1$ minimal models in \Sec{n1mm}, and we present new CFT candidates and interpret their properties in \Sec{hunt}.
We then perform a survey over possible central charges and spectral gaps in \Sec{survey}, where we identify a primal/dual gap between the dual $\Delta_{\rm gap} \le \frac{c}{6} + \frac{1}{3}$ bound and our primal constructions.
We conclude in \Sec{conclude}, leaving additional cross-checks and plots to the appendices.

\section{Review of the Modular Bootstrap}
\label{sec:review}

We start with a review of the 2d CFT modular bootstrap, in order to set notation and define the target for our ML-style optimization.

\subsection{Properties of Unitary 2d CFTs}

The basic information about a 2d CFT is its central charge $c \ge 0$ and its spectrum of Virasoro primary operators:
\begin{equation}
\{\Delta_a,j_a,d_a\},
\end{equation}
which is a set of dimensions $\Delta_a$, spins $j_a$, and degeneracies $d_a$.
We work exclusively with discrete spectra, such that the index $a$ is countably infinite.
Unitarity  imposes a constraint that the dimension of an operator is bounded below by its spin, and that the degeneracy is a non-negative integer:%
\footnote{A state with $d_a = 0$ does not actually contribute to the spectrum, but we allow it in our code implementation, since it is a consistent choice.}
\begin{equation}
    \label{eq:unitarity}
    \Delta_a \ge |j_a|, \qquad d_a \in \mathbb{N} = \{0,1,2,\ldots\}.
\end{equation}
We further focus on parity-invariant theories compactified on a Euclidean torus, which implies that states come in pairs with equal and opposite spin $j \leftrightarrow -j$. For this reason, we omit the negative spin states in the visualizations below, though they are included in the computations.
 We furthermore assume we can work with integer-valued spins:
\begin{equation}
    \label{eq:integer_spin}
    j_a \in \mathbb{Z} = \{0,\pm 1, \pm 2,\ldots\},
\end{equation}
which essentially means we are restricting our study to bosonic CFTs (or more generally, the bosonic subsectors of CFTs). 
To fully specify a CFT, we would also need to provide OPE coefficients, but those do not appear in the modular bootstrap equations.

Every discrete unitary 2d CFT has a unique vacuum operator:
\begin{equation}
(\Delta_{\rm vac}, j_{\rm vac},d_{\rm vac}) = (0,0,1).
\end{equation}
It is convenient to characterize the rest of the CFT spectrum in terms of its spectral gap and twist gap, respectively:%
\footnote{Due to the possibility of an accumulation point in twist, we define $t_{\text{gap}}$ as the infimum rather than the minimum of the twists of the non-vacuum Virasoro primaries. An accumulation point in twist always occurs, for instance, in any theory with a $u(1)$ symmetry \cite{Benjamin:2020swg}.}
\begin{equation}
    \Delta_{\rm gap} = \min_{a \not= {\rm vac}} \Delta_a, \qquad     t_{\rm gap} = \inf_{a \not= {\rm vac}} \big(\Delta_a - |j_a| \big).
\end{equation}
For discrete spectra, the spectral gap is strictly positive:
\begin{equation}
    \Delta_{\rm gap}>0.
\end{equation}
In our numerical studies, we focus on irrational CFTs that do not have any conserved currents beyond the stress tensor.
This is equivalent to imposing a strictly positive twist gap:
\begin{equation}
    \label{eq:twist_gap}
    t_{\rm gap}>0,
\end{equation}
which excludes rational CFTs with additional conserved currents ($\Delta_{\rm current} = |j_{\rm current}|$).
As we explain at the end of \Sec{partition_function} below, though, there is a continuous way to approach the $t_{\rm gap} \to 0$ limit, so our approach can in principle identify some, though not all, rational CFTs.
Since all CFTs with central charge less than 1 have been cataloged and are known to be rational, our irrational candidates will have:
\begin{equation}
    c \ge 1.
\end{equation}

An important constraint on the spectrum of CFT primaries is known as the Hellerman-Collier-Lin-Yin (HCLY) bound, which constrains the size of the spectral gap:
\begin{equation}
\label{eq:hellerman}
    \Delta_{\rm gap} \le \frac{c}{6} + \frac{1}{3} \quad \text{when }\quad1\leq c\leq4.
\end{equation}
The $\frac{c}{6}$ term appears in Hellerman's original paper~\cite{Hellerman:2009bu}, albeit with a constant factor of $0.473694978898$.
This constant was slightly refined in \Reference{Friedan:2013cba}, and the factor of $\frac{1}{3}$ was derived in the Collier-Lin-Yin paper~\cite{Collier:2016cls}.
Explicit modular-invariant partition functions that saturate the updated bound can be constructed using the extremal functional method (see e.g.~\Reference{Fitzpatrick:2023lvh} for the explicit spectra), albeit without the constraint of integer-valued degeneracy.
Assuming integer-valued degeneracy, it is possible to obtain percent level improvements of bound for $c \gtrsim 1$~\cite{Fitzpatrick:2023lvh}.
We will not impose \Eq{hellerman} in our studies, but rather use it as a diagnostic for the survey in \Sec{survey}.

\begin{figure}[p]
    \centering
    \subfloat[][]{
        \includegraphics[width=0.4\linewidth]{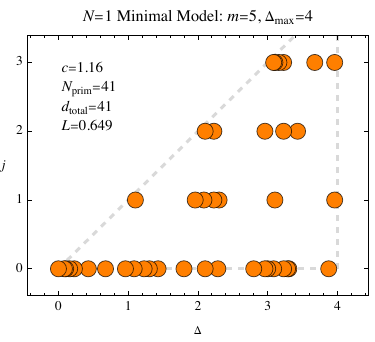}
        \label{fig:n1mm_spectrum_m5}    
    }\\
    \subfloat[][]{
        \includegraphics[width=0.4\linewidth]{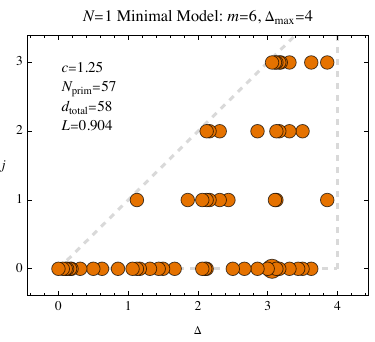}
    }$\qquad$
    \subfloat[][]{
        \includegraphics[width=0.4\linewidth]{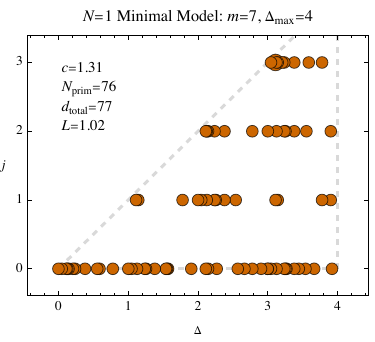}
    }\\
    \subfloat[][]{
        \includegraphics[width=0.4\linewidth]{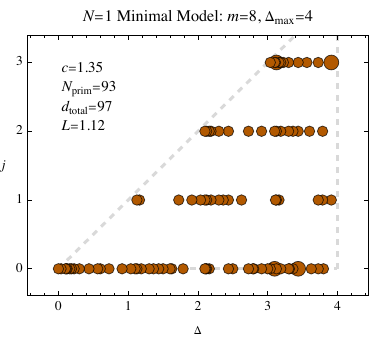}
    }$\qquad$
    \subfloat[][]{
        \includegraphics[width=0.4\linewidth]{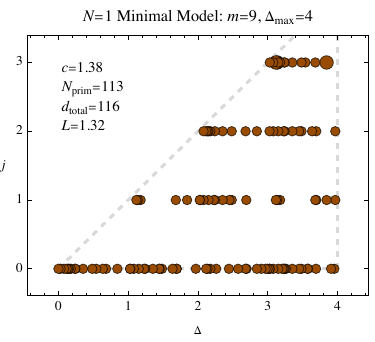}
    }
    \caption{Spectrum of Virasoro primary operators below $\Delta_{\rm max} = 4$ for the $N=1$ minimal model (N1MM) with (a) $m = 5$, (b) $m=6$, (c) $m=7$, (d) $m=8$, and (e) $m=9$.
    Here and in other spectrum plots below, the degeneracy of each primary is proportional to the area of the corresponding dot, with the vacuum at $(\Delta,j) = (0,0)$ having degeneracy one, and every operator with positive spin has a negative spin counterpart.
    These five N1MM examples will be used to benchmark the loss value corresponding to CFT-like behavior, with $m = 5$ in \Fig{n1mm_spectrum_m5} used as our default.
    The losses $L$ come from \Tab{n1mm_m_vs_r_table} below, where order 1 values are considered CFT-like.}
    \label{fig:n1mm_spectrum}   
\end{figure}

In \Fig{n1mm_spectrum}, we show example CFT spectra for the $N=1$ minimal models (N1MMs)  with $m \in \{5,6,7,8,9\}$~\cite{Friedan:1984rv}.
The parameter $m$ sets the central charge and spectral gap via:\footnote{For $m\in\{3,4\}$, the formula in \Eq{N1MMGapandC} does not apply and the gaps are given by $\{\frac{3}{80}, \frac{1}{24}\}$, respectively.}
\begin{equation}
c_{\rm N1MM} = \frac{3}{2} \left(1 - \frac{8}{m (m + 2)}\right), \qquad \Delta_{\rm gap}^{\rm N1MM} = \frac{3}{m(m+2)}.
\label{eq:N1MMGapandC}
\end{equation}
To derive the primary operator spectra, we took the known partition functions for these theories from \Reference{Kiritsis:1986rv}, and did a term by term matching to Virasoro characters via \Eq{partition_function} below.
The N1MM theories are rational CFTs with $t_{\rm gap}= 0$, so there are non-vacuum operators on the $\Delta = |j|$ line.
While the spectrum of operators is infinite, we have imposed a truncation to only show operators with $\Delta_a < \Delta_{\rm max}$.
In \Sec{n1mm}, these truncated N1MM theories will be used as benchmark models to validate our analysis procedure.

\begin{figure}
    \centering
    \subfloat[][]{
        \includegraphics[width=0.4\linewidth]{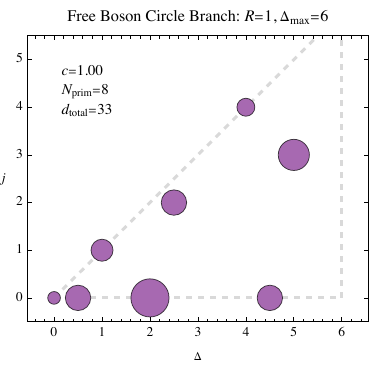}
    }$\qquad$
    \subfloat[][]{
        \includegraphics[width=0.4\linewidth]{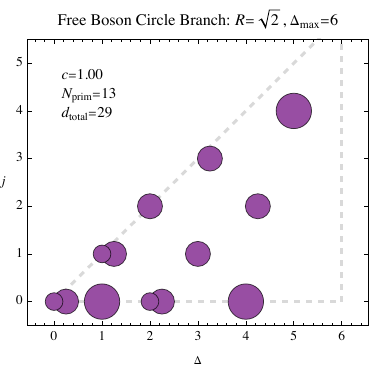}
        \label{fig:fbcb_spectrum_rsprt2}    
    }\\
    \subfloat[][]{
        \includegraphics[width=0.4\linewidth]{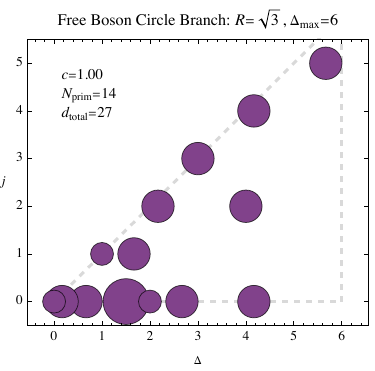}
    }$\qquad$
    \subfloat[][]{
        \includegraphics[width=0.4\linewidth]{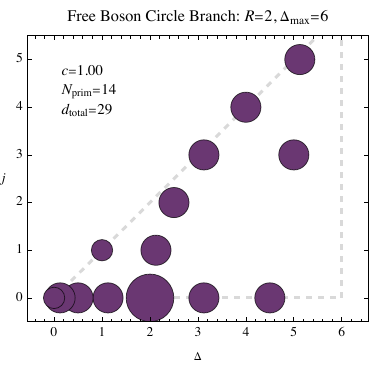}
    }
    \caption{Spectrum of Virasoro primary operators below $\Delta_{\rm max} = 6$ for free boson theories on the circle branch (FBCB) with radii (a) $R = 1$, (b) $R = \sqrt{2}$, (c) $R = \sqrt{3}$, and (d) $R = 2$.
    These four FBCB theories will be used as baselines to estimate the uncertainty associated with truncating the spectrum at $\Delta_{\rm max}$.
    To estimate uncertainties for the N1MM models in \Fig{n1mm_spectrum} with $\Delta_{\rm max} = 4$, we take these FBCB baselines to have $\Delta_{\rm max} = 6$.
    The spectrum of the FBCB theory with $R = \sqrt{2}$ in \Fig{fbcb_spectrum_rsprt2} is most similar visually to those of the N1MM models, and we will use it for our final truncation uncertainty estimate.
    }
    \label{fig:fbcb_spectrum}    
\end{figure}

In \Fig{fbcb_spectrum}, we show example CFT spectra for the free boson circle branch (FBCB) theories with $R \in \{1,\sqrt{2},\sqrt{3},2\}$.
The central charge and spectral gaps for these theories are:
\begin{equation}
    c_{\rm FBCB} = 1, \qquad \Delta_{\rm gap}^{\rm FBCB} = \frac{1}{2}\min\Big(R^2,\frac{1}{R^2} \Big),
\end{equation}
where $R = 1$ corresponds to the self-dual radius.
The spectra for these theories are well-known~\cite{Ginsparg:1987eb,Ginsparg:1988ui}, though we also performed a partition function matching for validation.
Since we have picked rational values of $R^2$ for these examples, they happen to all be rational CFTs, though irrational values of $R^2$ lead to irrational CFTs.
In \Sec{truncation}, these FBCB theories will be used to help estimate truncation uncertainties.

\subsection{Euclidean Torus Partition Function}
\label{sec:partition_function}

Given the central charge and primary spectrum of a CFT, we can uniquely determine its Euclidean torus partition function $Z(\tau)$.
The torus is parametrized by a complex number $\tau$, and it is sometimes helpful to explicitly separate out its real and imaginary components:
\begin{equation}
\label{eq:tau_re_im}
    \tau = \tau_1 + i \, \tau_2.
\end{equation}
Because we are working on a Euclidean torus, $\bar{\tau}$ is the complex conjugate of $\tau$:
\begin{equation}
    \bar{\tau} = \tau_1 - i \, \tau_2.
\end{equation}
It is convenient to define the squared-nome $q$ and its complex conjugate $\bar{q}$:%
\footnote{\label{footnote:claude3}In the mathematics literature, $q$ usually represents the nome $e^{\pi i \tau}$, which can be a source of confusion, especially for sycophantic chatbots.}
\begin{equation}
q = e^{2 \pi i \tau}, \qquad \bar{q} = e^{-2 \pi i \bar{\tau}}.
\end{equation}
While it is possible to define a more general torus partition function $Z(\tau,\bar{\tau})$ where $\tau$ and $\bar{\tau}$ are independent, we leave such studies to future work.

Each operator $(\Delta_a,j_a,d_a)$ in the spectrum has associated conformal weights:
\begin{equation}
 h_a = \frac{\Delta_a + j_a}{2}, \qquad \bar{h}_a = \frac{\Delta_a - j_a}{2},
\end{equation}
where $h$ and $\bar{h}$ are non-negative by \Eq{unitarity}.
For irrational CFTs, the twist gap in \Eq{twist_gap} implies that only the vacuum state has $h = 0$ or $\bar{h} = 0$.
The partition function depends on the Virasoro characters for each operator, with an important discontinuity at $h = 0$:
\begin{align}
    \chi_{h=0}(\tau) &= \frac{q^{- \frac{c-1}{24}}}{\eta(\tau)} (1-q), &
    \chi_{h>0}(\tau) &=  \frac{
    q^{h - \frac{c-1}{24}}
    }{\eta(\tau)}, \nonumber\\
    \overline{\chi}_{h=0}(\bar{\tau}) &= \frac{\bar{q}^{- \frac{c-1}{24}}}{\eta(-\bar{\tau})} (1-\bar{q}), &
    \overline{\chi}_{\bar{h}>0}(\bar{\tau}) &=  \frac{
    \bar{q}^{\bar{h} - \frac{c-1}{24}}
    }{\eta(-\bar{\tau})},
    \label{eq:virasoro_characters}
\end{align}
where $\eta(\tau)$ is the Dedekind eta function, and we have used the identity $\overline{\eta(\tau)} = \eta(-\overline{\tau})$.

Putting these ingredients together, the Euclidean torus partition function is
\begin{equation}
    \label{eq:partition_function}
    Z(\tau) = \sum_{a} d_a \, \chi_{h_a}(\tau) \, \overline{\chi}_{\bar{h}_a}(\bar{\tau}).
\end{equation}
Note that this sum over $a$ includes the vacuum operator and the positive/negative spin pairings implied by parity invariance.
For the numerical optimization, we truncate the spectrum at a maximum dimension, such that the sum over $a$ only involves operators with
\begin{equation}
    \label{eq:delta_max}
    \Delta_a < \Delta_{\rm max}.
\end{equation}
In our code implementation, we also find it convenient to work with the reduced partition function~\cite{Friedan:2013cba}:
\begin{equation}
    \label{eq:reduced_partition}
    Z_{\rm red}(\tau) = \sqrt{\tau_2} \, |\eta(\tau)|^2 \, Z(\tau),
\end{equation}
which allows us to minimize calls to the Dedekind eta function, but otherwise has no impact on the analysis.

Though we enforce a twist gap on the spectrum, some of the solutions we find have $t_{\rm gap}$ very close to zero.
While the Virasoro characters are discontinuous in the $t_{\rm gap} \to 0$ limit, the partition function smoothly interpolates from an irrational CFT to a rational one.
Consider an operator with spin $x$, twist $\epsilon$, and degeneracy 1:
\begin{equation}
    (\Delta,j,d) = (x+\epsilon,x,1).
\end{equation}
In the $\epsilon \to 0$ limit, the contribution to the partition function from this single operator can be mimicked by two operators with
\begin{equation}
    (\Delta,j,d) = (x,x,1), \qquad (\Delta,j,d) = (x+1,x-1,1).
\end{equation}
The first operator corresponds to a spin $x$ conserved current, while the second operator compensates for the $(1-q)$ factor in the first's character.  In other words, at the point when the twist gap closes, the representation splits into a short representation and a long representation, which are a conserved current and a second operator.
The degeneracy of this second operator is positive, so it still satisfies the unitarity constraint from \Eq{unitarity}.

\subsection{Constraints from Modular Invariance}

When acting on the Euclidean torus, conformal symmetry together with large gauge transformations imply modular invariance of the partition function:
\begin{equation}
    Z(\tau) = Z\left(\frac{a \, \tau + b}{c \, \tau + d}\right), 
\end{equation}
where $a$, $b$, $c$, and $d$ are integers satisfying $ad-bc=1$, i.e.~elements of $\text{SL}(2,\mathbb{Z})$.
The modular group is generated by $S$ and $T$ transformations:
\begin{equation}
\tau_T =\tau + 1, \qquad
    \tau_S = - \frac{1}{\tau} = \frac{- \tau_1 + i\, \tau_2}{\tau_1^2 + \tau_2^2}.
\end{equation}
Because we are working with integer-valued spins via \Eq{integer_spin}, $Z(\tau) = Z(\tau_T)$ by construction.

The modular bootstrap is then defined by the remaining non-trivial modular constraint from $S$ transformations:
\begin{equation}
    \label{eq:S_invariance}
    Z(\tau) = Z(\tau_S).
\end{equation}
Numerical solutions to \Eq{S_invariance} will be the target for ML-style optimization in \Sec{loss}.
Equivalently, we can work with the reduced partition function from \Eq{reduced_partition} and impose $Z_{\rm red}(\tau) = Z_{\rm red}(\tau_S)$, because the following factor is invariant to $S$ transformations:%
\footnote{This factor is \emph{not} invariant to $T$ transformations.  However, we have already imposed the integer spin relation, which imposes the $T$ invariance of the partition function.  Moreover, the full modular group $\text{SL}(2,\mathbb{Z})$ is generated by $S$ and $T$, so it is sufficient to also impose invariance under $S$, which we can do by working with the reduced partition function.}
\begin{equation}
\label{eq:s_invariant_factor}
\sqrt{\tau_2} \, |\eta(\tau)|^2.
\end{equation}
The $S$ invariance of \Eq{s_invariant_factor} will also feature in the deformation strategy of \Sec{deformation_strategy}.

The constraint in \Eq{S_invariance} holds for every choice of $\tau$.
To turn this into a numerical optimization target, we have to choose specific $\tau$ values.
In the dual bootstrap, it is common to take derivatives of the partition function, evaluated at the self-dual point $\tau = i$; we leave an exploration of that approach to future work.
As discussed further in \Sec{tau_samples}, we choose to sample $\tau$ uniformly in the fundamental domain according to a modular invariant measure.
As we will see, though, the choice of $\tau$ measure is a subdominant effect compared to choosing $\tau$-dependent uncertainties.

\section{Machine-Learning-Style Optimization}
\label{sec:ml}

We now translate the modular bootstrap program above into the language of ML.
We refer to this approach as ``ML-style,'' since it is similar to traditional numerical optimization, but with a stochastic sampling approach inspired by the ML literature.
Crucially, though, no neural networks are involved and no training data from known CFTs are needed, so in that sense, it is different from other approaches appearing under the banner of ``machine learning.''
A \mathematica\ implementation of this workflow is available from GitHub at \url{http://github.com/jdthaler/modular-bootstrap}.

\subsection{Optimizing the Partition Function}
\label{sec:partition_function_opt}

In our primal approach to the modular bootstrap, we directly optimize the partition function in \Eq{partition_function}.
While the partition function is parametrized by the central charge $c$ and primary spectrum $\{\Delta_a,j_a,d_a\}$, we only optimize over the dimensions $\{\Delta_a\}$:
\begin{equation}
    Z(\tau; c, \{\Delta_a,j_a,d_a\}) \Rightarrow Z(\tau; \{\Delta_a\}).
\end{equation}
That is, we fix $c$ and a set of spin/degeneracy pairs $\{j_a,d_a\}$, choose initial conditions for the set of dimensions $\{\Delta_a\}$, and then only optimize the $\Delta_a$ values.
For the survey in \Sec{survey}, we also fix the spectral gap $\Delta_{\rm gap}$ by setting the lowest spin-0 non-vacuum primary to have $\Delta_1 = \Delta_{\rm gap}$.

The reason we fix the central charge $c$ is twofold.
First, the truncation uncertainty estimate in \Sec{truncation} is strongly $c$ dependent, and if $c$ were allowed to float, we would have to do more computationally expensive updates.
Second, it turns out to be easier to find solutions to the truncated modular bootstrap when $c$ is small, since fewer operators are needed to satisfy modular invariance, so $c$ tends to drift towards 1 if left unconstrained.
For the spin and degeneracy, we fix their values for the practical reason that they are discrete integers, and we wanted to avoid mixed-integer optimization.%
\footnote{As a consistency check, we do one optimization campaign in \Sec{descend_gap} where we let $d_a$ float and take non-integer values.}
Without loss of generality, we set all operators to have degeneracy $d_a = 1$, since higher values of $d_a$ can be obtained by having multiple operators with the same spin and dimension.
The number of operators and their spins is an important choice we have to make when setting up the optimization.

To implement the constraint from \Eq{unitarity} that  $\Delta_a \ge |j_a|$, we map an unconstrained parameter $x_a$ to a finite interval $\Delta_a \in (\Delta_{\rm lower},\Delta_{\rm upper})$:
\begin{equation}
    \label{eq:map_to_gap}
    \Delta_a = g(x_a; \Delta_{\rm lower}, \Delta_{\rm upper}).
\end{equation}
For concreteness, we use the following function with sigmoid-like behavior:
\begin{equation}
    g(x; \Delta_{\rm lower}, \Delta_{\rm upper}) = \frac{\Delta_{\rm lower} + \Delta_{\rm upper} \, e^x}{1 + e^x}.
\end{equation}
We choose bounds on the interval of:
\begin{equation}
\Delta_{\rm lower} = \max(|j_a|,\Delta_{\rm gap}), \qquad \Delta_{\rm upper} = 2 \, \Delta_{\max} - \Delta_{\rm lower},
\end{equation}
such that $x_a = 0$ corresponds to $\Delta_a = \Delta_{\max}$.
We initialize all of the $x_a$ values to be negative, so all operators start off with dimension below the truncation scale.
During optimization, we let the operator dimensions cross the $\Delta_{\max}$ threshold without restriction, in order to avoid the problem of ``vanishing gradients'' that can arise in ML optimization.
Since we do not know how many operators will be required below $\Delta_{\max}$ to satisfy modular invariance, we purposely start with more operators than we think we need, and allow them to drift above $x_a = 0$ if they are not needed.
When we report the final loss, we truncate the spectrum to only include operators with $x_a < 0$ (i.e.~$\Delta_a < \Delta_{\rm max}$), so that we are working with a cleanly truncated partition function.

\subsection{Choice of Loss Function}
\label{sec:loss}

The naive translation of the modular bootstrap equation in \Eq{S_invariance} into a ML-style loss function is:
\begin{equation}
\label{eq:naive_loss}
L_{\rm naive} = \frac{1}{|\mathcal{T}|}\sum_{\tau \in \mathcal{T}} \big(Z(\tau) - Z(\tau_S)\big)^2,
\end{equation}
where $\mathcal{T}$ is a suitable set of $\tau$ values with cardinality $|\mathcal{T}|$.
This loss is of the mean squared error (MSE) form, which is standard in the ML literature.
Certainly, if modular invariance holds exactly, then this MSE loss is zero, as desired.

The loss in \Eq{naive_loss} has a number of undesirable features, though.
The first issue is that this loss is exponentially sensitive to the choice of truncation scale $\Delta_{\rm max}$ in \Eq{delta_max}.
For a real CFT, the loss decreases as $\Delta_{\rm max}$ increases, since the truncated partition function gets closer to the true partition function.
Because the product of Virasoro characters involves factors of $|q|^{\Delta}$, though, the loss changes exponentially with $\Delta_{\rm max}$ when $\tau_2$ is sufficiently large.
The second issue, closely related to the first, is sensitivity to the precise set of $\tau$ values that contribute to the loss.
While some sensitivity to $\mathcal{T}$ is inevitable, the loss is highly dependent on the maximum value of $\tau_2$ selected, because of the same $|q|^{\Delta}$ factors mentioned above.
The third issue is the difficulty to calibrate the loss.
Because of the $|q|^{c/12}$ terms in the partition function, known CFTs with different central charges have different losses, even when they are truncated to the same scale $\Delta_{\rm max}$ and evaluated on the same set $\mathcal{T}$.
Since we are hunting for new CFT candidates, it is difficult to know what values of the loss are close enough to zero to declare them to be sufficiently CFT-like.

All of these issues can be resolved by using an improved $\chi^2$-style loss that is normalized by a $\tau$-dependent uncertainty:
\begin{equation}
\label{eq:improved_loss}
L_{\rm improved} = \frac{1}{|\mathcal{T}|}\sum_{\tau \in \mathcal{T}} \left(\frac{Z(\tau) - Z(\tau_S)}{\sigma_{\rm trunc}(\tau)} \right)^2.
\end{equation}
This loss deemphasizes contributions from $\tau$ values that are associated with larger uncertainties.
The dominant uncertainty comes from the truncation scale $\Delta_{\rm max}$, and we provide a concrete strategy to extract $\sigma_{\rm trunc}(\tau)$ in \Sec{truncation}.
Including this uncertainty turns out to automatically mitigate the $\mathcal{T}$ sensitivity and calibration issues, since $\sigma_{\rm trunc}(\tau)$ compensates for the exponential sensitivity to $\Delta_{\rm max}$.

As shown later in \Sec{n1mm}, the individual terms in this improved loss follow a $\chi^2$ distribution, such that order $1$ values of the loss correspond to CFT-like behavior.
Perhaps counterintuitively, as $\Delta_{\rm max}$ increases, the improved loss does \emph{not} decrease.
Rather, both the numerator and denominator of \Eq{improved_loss} shrink in a correlated way, such that order $1$ loss values still indicate CFT-like behavior, even in the $\Delta_{\rm max} \to \infty$ limit.
This is a feature, not a bug, which allows us to set an unambiguous accuracy target for any choice of $\Delta_{\rm max}$.

\subsection{Training, Testing, and Reporting Samples}
\label{sec:tau_samples}

A key ingredient in any ML problem is a set of training data.
In our case, the training data are samples of $\tau$ values from (half of) the fundamental domain, which are used to evaluate the loss in \Eq{improved_loss}.
In typical ML applications, the training data come from a fixed data set.
Here, we can select as many $\tau$ values as desired, which mitigates the problem of overfitting.
Translating from the ML jargon, sampling the training data is equivalent to performing Monte Carlo integration over $\tau$.

Because we only consider parity-invariant CFTs, we can restrict the sampling to $\tau_1 > 0$.
We select $\tau$ values uniformly in the fundamental domain according to the modular invariant measure, albeit with a cut on $\tau_2^{\rm max}$ for numerical stability:
\begin{equation}
\label{eq:tau_measure}
    \int_0^{1/2} d \tau_1 \, \int_0^{\tau_2^{\rm max}} \frac{d\tau_2}{\tau_2^2} \, \Theta(|\tau| - 1),
\end{equation}
where $\Theta$ is the Heaviside theta function.
To handle the non-trivial $\frac{1}{\tau_2^2}$ measure, we sample an auxiliary variable $u$ uniformly between $(0,u_{\rm max})$ and convert it to $\tau_2$ via:
\begin{equation}
    \tau_2 = \frac{\sqrt{3}}{2}\frac{1}{1-u},
\end{equation}
where $u_{\rm max}$ is chosen to match $\tau_2^{\rm max}$, and the factor of $\frac{\sqrt{3}}{2}$ corresponds to the smallest value of $\tau_2$ consistent with the $|\tau| > 1$ constraint.
By default, we set:
\begin{equation}
\tau_2^{\rm max} = \Delta_{\rm max} + 2,
\end{equation}
though we consider alternative choices in \Fig{n1mm_truncation_scans} below.
As we will see, the dependence of the loss on $\tau_2^{\rm max}$ is relatively mild, as long as we use the improved version from \Eq{improved_loss}.

\begin{figure}
    \centering
    \subfloat[][]{
    \label{fig:tau_training}
    \includegraphics[width=0.29\linewidth]{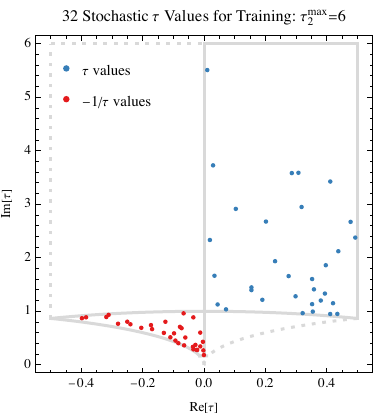}
    }
    $\quad$
    \subfloat[][]{
    \label{fig:tau_testing}
    \includegraphics[width=0.29\linewidth]{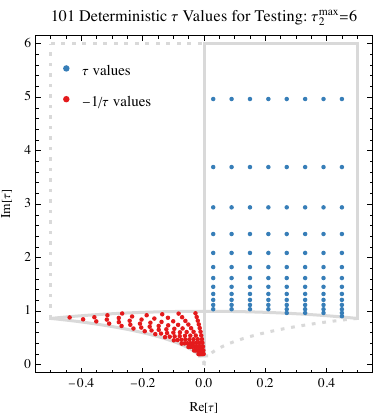}
    }
    $\quad$
    \subfloat[][]{
    \label{fig:tau_reporting}
    \includegraphics[width=0.29\linewidth]{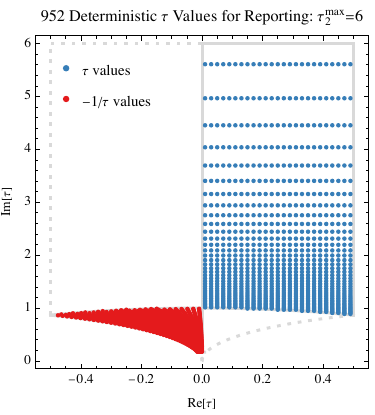}
    }
    \caption{Values of $\tau = \tau_1 + i \, \tau_2$ in the fundamental domain used for (a) training, (b) testing, and (c) final reporting, along with their $S$-transformed counterparts.
    For numerical stability, we restrict $\tau_2 < \tau_2^{\rm max}$.
    Because we restrict to parity-preserving CFTs, we only need to consider $\tau_1 > 0$.
    During training, new $\tau$ values are chosen stochastically for each iteration.
    During testing and reporting, the same fixed grid of $\tau$ values are used, with higher fidelity during the reporting stage.
    Because the ML hyperparameters are fixed throughout the training process, we do not need a validation set of $\tau$ values.}
    \label{fig:taus}
\end{figure}

There are three distinct phases in the ML pipeline, with different $\tau$ ranges shown in \Fig{taus}.
During training, when we compute derivatives of the loss function, we use stochastic sampling to avoid the possibility of memorizing the training data.
As shown in \Fig{tau_training}, we use the following default number of samples per iteration:
\begin{equation}
N_\tau^{\rm train} = 32.
\end{equation}
This number was chosen to be twice the number of singular values used for the default \sven\ algorithm, as discussed more in \Sec{sven}.
With too few training samples, the computation of the gradients is noisy.
With too many training samples, not only do the computational costs go up, but it becomes more likely for the algorithm to get stuck in local loss minima.
As mentioned in \Sec{partition_function_opt}, we do not impose the $\Delta_a < \Delta_{\rm max}$ constraint during training, to avoid the problem of vanishing gradients.

During testing, when we evaluate the loss after performing an update, we use a deterministic grid of $\tau$ values that follows \Eq{tau_measure}.
This is shown in \Fig{tau_testing} for the case of $\tau_2^{\rm max} = 6$:
\begin{equation}
N_\tau^{\rm test} = 101.
\end{equation}
During training, the test loss might not monotonically decrease.
For the hunt in \Sec{hunt}, we ignore this effect and simply take the model at the end of training, even if there were a higher-quality intermediate result.
Thus, $N_\tau^{\rm test}$ only affects the visual appearance of \Fig{campaign_loss_vs_iter} below, with larger $N_\tau^{\rm test}$ values yielding smoother behavior.
For the survey in \Sec{survey}, where it is computationally more expensive to find good candidates, we use the testing data set to select the best candidate spectrum for further analysis.
In this context, the value of $N_\tau^{\rm test}$ affects what spectra are considered ``best.''
We decided not to impose $\Delta_a < \Delta_{\rm max}$ during the testing stage, though it would be straightforward to do so.%
\footnote{It would also be possible for us to go back through the training record and recompute the test loss with the $\Delta_{\rm max}$ cut, since the value of the test loss has no impact on the training dynamics.}

Finally, when reporting final results, we do impose the $\Delta_a < \Delta_{\rm max}$ constraint and also use a finer $\tau$ grid, as shown in \Fig{tau_reporting} for $\tau_2^{\rm max} = 6$:
\begin{equation}
N_\tau^{\rm report} = 952.
\end{equation}
The difference between using $N_\tau^{\rm test}$ and $N_\tau^{\rm report}$ to compute the loss is typically larger than the 5--10\% level expected from the scaling of statistical fluctuations, due to the impact of $\Delta_{\rm max}$.
By using the reported loss, we are guarding against false positives at the expense of possibly throwing out spectra that might be close to CFT-like with further optimization.
This is important to keep in mind for the survey in \Sec{survey}, where we identify a gap between the dual bound and primal examples, though the gap still persists even if we use the smaller testing (or even training) loss.

\subsection{Gradient Descent versus Singular Value Descent}
\label{sec:sven}

The final ML ingredient we need is a choice of optimizer.
While we have not done an extensive comparison of different optimization algorithms, the recent \sven\ algorithm~\cite{Bright-Thonney:2026tlr} is well matched to this problem.
Because \sven\ allows multiple parameter directions to be explored simultaneously, it requires fewer optimization steps to find a good minimum.
On the other hand, \sven\ requires storing and manipulating more information than in standard gradient descent, so there is a limit to the number of parameters that can be simultaneously optimized.
This is not an issue for the present study, though, because we are only ever optimizing over $\mathcal{O}(100)$ parameters, and computing the gradients is the main computational bottleneck.

A complete motivation for and description of the \sven\ algorithm appears in \Reference{Bright-Thonney:2026tlr}.
Here, we present an intuitive way to understand how it works.
Consider a generic loss function $L$ that takes the form of a sum of squares:
\begin{equation}
\label{eq:generic_loss}
    L = \sum_i \big(R_i(p_a) \big)^2,
\end{equation}
where the residuals $R_i$ depend on parameters $p_a$.
For the improved loss in \Eq{improved_loss}, the sum over $i$ runs over different $\tau$ values, the residuals involve the (uncertainty normalized) partition function difference, and the parameters $p_a$ are the operator dimensions $\Delta_a$.
While \sven\ works for any choice of loss function, it is easiest to explain in this least squares setting.

Imagine Taylor expanding the residuals around the initial values of the parameters:
\begin{equation}
\label{eq:sven_expansion}
    p_a = p_a^{(0)} + \delta_a, \qquad 
    R_i = R_i^{(0)} + \sum_a M^{(0)}_{ia} \, \delta_a  + \cdots,
\end{equation}
where $R_i^{(0)} \equiv R_i\big(p_a^{(0)}\big)$ and the Jacobian matrix is:
\begin{equation}
    M^{(0)}_{ia} = \left.\frac{\partial R_i}{\partial p_a}\right|_{p_a = p_a^{(0)}}.
\end{equation}
The ellipsis in \Eq{sven_expansion} represents quadratic and higher terms in $\delta_a$.
If these additional terms were absent, then \Eq{generic_loss} would be the objective function for standard linear least squares regression, whose unique solution (after possible $L^2$-regularization to remove degeneracies) is:
\begin{equation}
    \delta_a = - \sum_i (M^+)^{(0)}_{ai} \, R_i^{(0)},
\end{equation}
where $M^+$ is the Moore--Penrose pseudoinverse~\cite{moore1920reciprocal,bjerhammar1951application,penrose1955generalized} of the matrix $M$.

When non-linear terms are present, we cannot immediately jump to the exact solution, but we can perform iterative updates, similar in spirit to the well-known Gauss--Newton algorithm.
A key difference, though, is that Gauss--Newton assumes that there are more residuals than parameters (the underparametrized regime), whereas ML applications tend to work with more parameters than residuals (the overparametrized regime).
Nevertheless, the pseudoinverse is equally applicable in either regime.
A downside to using the pseudoinverse, though, is that it can be highly noisy if there are approximate flat directions in the loss landscape.

The key insight of \sven\ is to perform a singular value decomposition of $M$ and only keep the top singular values that are less affected by noise.
This leads to a powerful yet stable optimizer, with the following  parameter update rule at iteration $s$:
\begin{equation}
    p_a^{(s)} = p_a^{(s-1)} - \eta \, \sum_i (M_{k_\text{SV}}^+)^{(s-1)}_{ai} \, R_i^{(s-1)},
\end{equation}
where the superscripts lift the ${}^{(0)}$ notation from \Eq{sven_expansion} to iteration number in the obvious way, $\eta \in (0,1)$ is an adjustable learning rate, and the $k_{\rm SV}$ subscript indicates truncating the matrix $M$ to keep only the highest $k$ singular values.
This is the idea behind the \sven\ algorithm, which works well on a variety of problems.
For our primal modular bootstrap, we take a default value of
\begin{equation}
\label{eq:sven_default}
k_{\rm SV} = 16, \qquad \eta = 0.1,
\end{equation}
though we will introduce one additional optimization trick below to help find better solutions.

It is instructive to compare \sven\ to standard gradient descent:
\begin{equation}
    p_a^{(s)} = p_a^{(s-1)} - \tilde{\eta} \, G_a^{(s-1)}, \qquad G_a^{(s-1)} = \sum_i (M^T)^{(s-1)}_{ai} R_i^{(s-1)}.
\end{equation}
Here, $G_a$ is (minus half of) the gradient of the loss, and $\tilde{\eta}$ is the standard learning rate.
To make gradient descent behave a bit more like \sven, we can rescale the learning rate by the Polyak step size~\cite{polyak1969minimization}:
\begin{equation}
    \tilde{\eta} = \eta \,  \frac{\sum_i R_i^{(s-1)}R_i^{(s-1)}}{\sum_a G_a^{(s-1)} G_a^{(s-1)}},
\end{equation}
which yields an effective learning rate $\eta$ in the same $(0,1)$ range as \sven.
Both gradient descent and \sven\ depend on the same Jacobian matrix $M$, but gradient descent is based on the transpose of $M$, such that one moves faster in directions with larger gradients.
By contrast, \sven\ is based on the pseudoinverse of $M$, such that one moves slower in directions with larger singular values.
This behavior might seem counterintuitive, since to decrease the loss, you might think you should go downhill as fast as possible.
In the case of hierarchical loss landscapes, though, you actually want to move a short distance in steep directions and longer distance in flatter directions.%
\footnote{If you are skiing in a half-pipe, going only in the steep direction is great for doing tricks, but you still have to move parallel to the walls if you want to get to the end of the run.  In the context of \sven, these two directions corresponds to two different singular values.}
Indeed, a limiting case of \sven\ is linear least squares regression, where you get the exact solution in one step with $\eta = 1$.
To the extent that the residuals are approximately linear, it makes sense to move fast in flat-ish directions that gradient descent would never want to explore.

In this way, \sven\ is able to effectively explore $k_{\rm SV}$ directions simultaneously in one iteration.
This is similar in spirit to conjugate gradient methods, which try to move in directions orthogonal to previous steps.
The main challenge is choosing the right value of $k_{\rm SV}$.
If $k_{\rm SV}$ is too small, then \sven\ does not explore enough parameter directions to find a good minimum.
If $k_{\rm SV}$ is too large, then noise from the pseudoinverse impedes learning.
The additional optimization trick we introduce is to adaptively choose the number of singular values up to a maximum $k_{\rm SV}$, but allowing for fewer singular values if needed.
Specifically, in each iteration we keep adding singular value directions up until the point when the training loss becomes higher than the starting loss or when we reach the maximum $k_{\rm SV}$.
This encourages \sven\ to move far in parameter space, but still in directions of decreasing training loss.

For future work, it would be interesting to perform a more detailed comparison of different optimization algorithms.
As discussed more below, the minima found by \sven\ appear to be part of a continuous space of (truncated) modular bootstrap solutions.
It would be interesting to see whether alternative optimization algorithms might reveal qualitatively different solutions, including isolated regions of parameter space that are completely inaccessible to \sven.

\subsection{Estimating Parameter Uncertainties}
\label{sec:parameter_uncertainties}

Once we have found a suitable minimum of the loss, we want to characterize the corresponding candidate CFT.
In particular, we would like to quantify how accurately we have estimated the values of the operator dimensions.
In our case, the parameters are highly correlated, so we need the complete covariance matrix to characterize the parameter uncertainties.

Treating the generic loss in \Eq{generic_loss} as if it were a $\chi^2$ test statistic, the standard inverse covariance matrix for the parameters $p_a$ is:
\begin{equation}
\left(\Sigma^{-1}\right)_{ab} = \frac{1}{2} \left.\frac{\partial^2{L}}{\partial p_a \, \partial p_b} \right|_{\widehat{p}},
\end{equation}
where $\widehat{p}$ indicates the parameter values at the loss minimum.
Because of numerical artifacts and incomplete optimization, this expression might not actually yield a positive semi-definite covariance matrix.
Therefore, we follow a standard trick and estimate the second derivatives in terms of products of first derivatives:
\begin{equation}
\label{eq:covariance_approx}
\left(\Sigma^{-1}\right)_{ab} = \left. \sum_i \left(\frac{\partial R_i}{\partial p_a} \, \frac{\partial R_i}{\partial p_b} + R_{i} \, \frac{\partial^2 R_i}{\partial p_a \, \partial p_b} \right) \right|_{\hat{p}} \approx \big(\widehat{M}^T \widehat{M}\big)_{ab}, \qquad \widehat{M}_{ia} = \left. \frac{\partial R_i}{\partial p_a} \right|_{\hat{p}}.
\end{equation}
This approximation works if the residuals $R_i$ are sufficiently small (and exact if the residuals are zero), and it is guaranteed to yield a healthy positive semi-definite covariance matrix, since $\widehat{M}^T \widehat{M}$ cannot have negative eigenvalues.%
\footnote{Crucially, $\widehat{M}_{ia}$ involves derivatives on the residuals, which are typically large.  Acting on the individual loss terms $L_i = R_i^2$, the derivatives $\partial L_i / \partial p_a = 2 R_i \widehat{M}_{ia}$ (with no sum on $i$) are typically small when the residual is small.}
Even though we only optimize over the dimensions $\{\Delta_a\}$, we include the central charge $c$ when estimating parameter uncertainties.

In the studies in \Sec{n1mm_para_uncertainties} and \Sec{hunt}, we find evidence for a continuous space of solutions.
Specifically, the inverse covariance matrix is rank deficient, so there are correlated parameter directions one can traverse that barely change the loss.
To help characterize this space, we find it informative to compute partial covariance matrices where we only vary a few parameters at a time.
This corresponds to looking at inverses of submatrices of $\Sigma^{-1}$, which is crucially different from looking at submatrices of $\Sigma$.

\section{Quantifying the Impact of Truncation}
\label{sec:truncation}

Because we are looking for numerical solutions to the modular bootstrap equations, we have to carefully treat the issue of uncertainties.
The dominant uncertainty is from truncating the spectrum to only include primaries with $\Delta < \Delta_{\rm max}$.
To estimate this uncertainty, we take a known CFT with baseline central charge $c_0$ and deform its partition function to match the asymptotic density of states of a CFT with $c > c_0$.
We then impose the $\Delta_{\rm max}$ truncation on the deformed partition function, which will yield a $\tau$-dependent estimate of the truncation uncertainty that can be used in the improved loss from \Eq{improved_loss}.
In \App{alpha_truncation}, we explore an alternative deformation strategy with similar performance.

\subsection{Deformation Strategy}
\label{sec:deformation_strategy}

Consider an exact CFT with central charge $c_0$ and partition function:
\begin{equation}
\label{eq:exact_CFT}
Z_{\rm exact}(\tau; c_0).
\end{equation}
Expanding around the $q \to 0$ limit, this partition function must behave like:
\begin{equation}
    Z_{\rm exact}(\tau;c_0) = |q|^{-\frac{c_0}{12}} \big(1 + \ldots \big),
\end{equation}
where the overall scaling is set by the vacuum character and the dots correspond to terms that scale as positive powers of $q$ and $\bar{q}$.
This partition function is modular invariant by assumption:
\begin{equation}
    Z_{\rm exact}(\tau; c_0) = Z_{\rm exact}(\tau_S; c_0).
\end{equation}

To model the partition function of a CFT with $c > c_0$, we apply the following deformation:
\begin{equation}
\label{eq:generic_deformation}
    Z_{\rm deformed}(\tau; c) = \left(\sqrt{\tau_2} \, |\eta(\tau)|^2  \right)^{c_0-c} Z_{\rm exact}(\tau; c_0),
\end{equation}
again using the notation $\tau_2 = \text{Im}(\tau)$ from \Eq{tau_re_im}.
A similar approach was pursued in \Reference{Benjamin:2021ygh} to define the primary partition function.%
\footnote{\label{footnote:claude2}Even though two of the authors of \Reference{Benjamin:2021ygh} are co-authors here, it was Claude Opus 4.1 \sout{who} that first identified the relevance of this approach for truncation uncertainty characterization.}
As previewed in \Eq{s_invariant_factor}, $\sqrt{\tau_2} \, |\eta(\tau)|^2$ is modular invariant, so this deformed partition function is as well.
Less obvious is that this partition function behaves like a CFT with central charge $c$.
To see this, we can expand \Eq{s_invariant_factor} around the $q \to 0$ limit:
\begin{equation}
\sqrt{\tau_2} \, |\eta (\tau)|^2 =  |q|^{\frac{1}{12}} \, \sqrt{\tfrac{1}{\pi}\log \tfrac{1}{|q|}} \, \big(1 + \ldots \big).
\end{equation}
Therefore, up to logarithmic corrections, the deformed partition function in \Eq{generic_deformation} has the vacuum behavior of a CFT with central charge $c$, as desired:
\begin{equation}
\label{eq:deformed_cusp_limit}
    Z_{\rm deformed}(\tau; c) = |q|^{-\frac{c}{12}}  \left(\tfrac{1}{\pi}\log \tfrac{1}{|q|} \right)^{c_0-c} \big(1+ \ldots \big).
\end{equation}
As will be more clear from the discussion below, this deformation strategy only works if $c > c_0$.

\subsection{Choice of Baseline CFT}

Since we are interested in studying candidate CFTs with $c \gtrsim 1$, we take a baseline CFT with $c_0 = 1$.
Any choice with a partition function that can be reliably computed will work for our purposes.
In practice, we choose our baseline CFT to be one of the FBCB theories shown in \Fig{fbcb_spectrum} with its own truncation scale $\Delta_{\rm max}^{\rm FBCB}$ taken to be larger than the $\Delta_{\rm max}$ of interest.

Our default baseline theory is an FBCB with $R = \sqrt{2}$.
While we take $\Delta_{\rm max}^{\rm FBCB} = \Delta_{\rm max} + 2$ in practice, it is instructive to consider the $\Delta_{\rm max}^{\rm FBCB} \to \infty$ limit where the partition function for this baseline CFT can be written in closed form:
\begin{equation}
\label{eq:baseline_cft}
    Z^{\rm baseline}_{\rm exact}(\tau) = \frac{|\Theta_2(\tau)|^2 + |\Theta_3(\tau)|^2 + |\Theta_4(\tau)|^2}{2 \, |\eta(
    \tau)|^2},
\end{equation}
where $\Theta_n(\tau)$ is a Jacobi theta function.
Expanding this partition function in the $q \to 0$ limit,
\begin{equation}
Z^{\rm baseline}_{\rm exact}(\tau) = |q|^{-\frac{c_0}{12}} \left(1 + 2 |q|^{\frac{1}{4}} + 4 |q| + (q + \bar{q})    + \ldots \right),
\end{equation}
we see that the first term is the expected vacuum contribution for a $c_0 = 1$ CFT.

The corresponding deformed partition function is:
\begin{equation}
\label{eq:specific_deformed_partition}
Z^{\rm baseline}_{\rm deformed}(\tau; c) = \left(\sqrt{\tau_2} \, |\eta(\tau)|^2  \right)^{1-c} \, \frac{|\Theta_2(\tau)|^2 + |\Theta_3(\tau)|^2 + |\Theta_4(\tau)|^2}{2\, |\eta(\tau)|^2}.
\end{equation}
In the $q \to 0$ limit, this behaves as:
\begin{equation}
Z^{\rm baseline}_{\rm deformed}(\tau; c) = |q|^{-\frac{c}{12}} \left(\tfrac{1}{\pi}\log \tfrac{1}{|q|} \right)^{1-c} \left(1 + 2 |q|^{\frac{1}{4}} + 4 |q| + c\, (q + \bar{q})    + \ldots \right).
\label{eq:q_expanded_deformed}
\end{equation}
In addition to the logarithmic factor, we see that the coefficient of the $(q + \bar{q})$ term is no longer an integer for generic $c$.
Therefore, while the deformed partition function is modular invariant, it cannot correspond to a true CFT.
Nevertheless, this deformation is sufficient for estimating the truncation uncertainties on a candidate CFT.

\subsection{Imposing the Truncation}

To get an estimate for the truncation uncertainty, we truncate (an estimate of) \Eq{specific_deformed_partition} at a given $\Delta_{\rm max}$.
Roughly speaking, this involves performing an inverse Laplace transform to go from $\tau_2$ space to $\Delta$ space, multiplying by the constraint
\begin{equation}
\label{eq:delta_max_cut_on_h}
    \Theta\big(\Delta_{\rm max} - \Delta \big),
\end{equation}
and then doing a final Laplace transform to get back to $\tau_2$ space.

Consider expanding the exact CFT partition function in \Eq{exact_CFT} with central charge $c_0$ as a series in $q$ and $\bar{q}$:
\begin{equation}
\label{eq:exact_k_kbar_expansion}
Z_{\rm exact}(\tau; c_0) = \frac{|q|^{\frac{1-c_0}{12}}}{|\eta(\tau)|^2} \sum_{h,\bar{h}} a_{h,\bar{h}} \, q^h \,  \bar{q}^{\bar{h}},
\end{equation}
where $h$ and $\bar{h}$ are discrete non-negative values.
Note that the coefficients $a_{h,\bar{h}}$ need not be positive, because of the $(1-q)$ factor in the vacuum Virasoro character $\chi_0$, but we continue to use the $h$ notation for simplicity.
For the baseline CFT in \Eq{baseline_cft}, extracting these $a_{h,\bar{h}}$ coefficients is as simple as expanding the Jacobi theta functions.
Indeed, doing such an expansion is effectively the way we made the spectrum plots in \Fig{fbcb_spectrum}.

Now consider expanding a true CFT with central charge $c$ as a series in $q$ and $\bar{q}$:
\begin{equation}
 Z_{\rm true}(\tau; c) = \frac{|q|^{\frac{1-c}{12}}}{|\eta(\tau)|^2} \sum_{h,\bar{h}} b_{h,\bar{h}} \, q^h \,  \bar{q}^{\bar{h}}.
\end{equation}
We want to know how much this partition function would change if we truncated the spectrum of primaries to only include those with $\Delta < \Delta_{\rm max}$.
Because of the possible appearance of vacuum characters, imposing a maximum dimension is not exactly the same as imposing $h + \bar{h} \leq \Delta_{\rm max}$, but we will ignore this subtlety in our analysis.
We therefore seek an estimate of
\begin{equation}
     Z^{\rm trunc}_{\rm true}(\tau; c;\Delta_{\rm max}) = \frac{|q|^{\frac{1-c}{12}}}{|\eta(\tau)|^2} \sum_{h,\bar{h}} b_{h,\bar{h}} \, q^h \,  \bar{q}^{\bar{h}} \, 
     \Theta(\Delta_{\rm max} - h - \bar{h}).
\end{equation}

Since the deformed partition function in \Eq{generic_deformation} does not correspond to a true CFT, the equivalent expansion in $h$ and $\bar{h}$ involves a density of states:
\begin{equation}
\label{eq:deformed_expansion}
 Z_{\rm deformed}(\tau; c) = \frac{|q|^{\frac{1-c}{12}}}{|\eta(\tau)|^2}\sum_j \int d \Delta \,  \rho_j(\Delta) \, q^{\frac{\Delta+j}{2}} \,  \bar{q}^{\frac{\Delta-j}{2}},
\end{equation}
where $j$ are non-negative integers.
Note that the spectrum in $j = h - \bar{h}$ is still discrete because the deformation in \Eq{generic_deformation} is real valued, but the spectrum in $\Delta = h + \bar{h}$ will be continuous.
Once we have determined $\rho_j(\Delta)$, we can impose the constraint on $\Delta_{\rm max}$ via \Eq{delta_max_cut_on_h} and then perform the integral over $\Delta$ to estimate the truncated partition function:
\begin{equation}
\label{eq:deformed_trunc}
     Z^{\rm trunc}_{\rm deformed}(\tau; c; \Delta_{\rm max}) = \frac{|q|^{\frac{1-c}{12}}}{|\eta(\tau)|^2} \sum_j \int d \Delta \,  \rho_j(\Delta) \, \Theta(\Delta_{\rm max} - \Delta) \, q^{\frac{\Delta+j}{2}} \,  \bar{q}^{\frac{\Delta-j}{2}}.
\end{equation}

To see what this procedure looks like in practice, the deformed partition function in \Eq{generic_deformation} can be written in terms of the exact partition function in \Eq{exact_k_kbar_expansion} as: 
\begin{align}
    Z_{\rm deformed}(\tau; c) &= \frac{|q|^{\frac{1-c}{12}}}{|\eta(\tau)|^2}
    \, \tau_2^{\frac{c_0 - c}{2}}
    \left( |\eta(\tau)|^2 \, |q|^{-\frac{1}{12}}\right)^{c_0 - c} \sum_{h, \bar{h}} a_{h,\bar{h}} \, q^h \,  \bar{q}^{\bar{h}} \nonumber \\
    &= \frac{|q|^{\frac{1-c}{12}}}{|\eta(\tau)|^2} \, \tau_2^{\frac{c_0 - c}{2}} \sum_{h, \bar{h}}  \widetilde{a}_{h,\bar{h}} \, q^h \,  \bar{q}^{\bar{h}}.
    \label{eq:deformed_via_coeffs}
\end{align}
In the second line, we have used the fact that $|\eta(\tau)|^2 \, |q|^{-\frac{1}{12}}$ has an expansion in non-negative powers of $q$ and $\bar{q}$, which allows us to define modified coefficients $\widetilde{a}_{h,\bar{h}}$ that generally depend on $c - c_0$.
It is convenient to rewrite this sum in terms of $j = h - \bar{h}$ and $\Delta = h + \bar{h}$:
\begin{equation}
    \tau_2^{\frac{c_0 - c}{2}} \sum_{h, \bar{h}}  \widetilde{a}_{h,\bar{h}} \, q^h \,  \bar{q}^{\bar{h}} = \sum_{j,\Delta} \widetilde{a}_{\frac{\Delta+j}{2},\frac{\Delta-j}{2}} \, e^{2\pi i \,  j \, \tau_1} \left( \tau_2^{\frac{c_0 - c}{2}} e^{-2\pi \, \Delta \, \tau_2 } \right).
\end{equation}
This is almost of the desired form, except we need to convert the term in parentheses into a density of states in order to use \Eq{deformed_trunc}.
This can be accomplished using the inverse Laplace transform:
\begin{equation}
\label{eq:inverse_Laplace_result}
\tau_2^{\frac{c_0 - c}{2}} e^{-2\pi \, \Delta \, \tau_2} = \frac{(2 \pi)^{\frac{c-c_0}{2}}}{\Gamma(\tfrac{c-c_0}{2})} \int d\widetilde{\Delta} \, \frac{\Theta(\widetilde{\Delta} - \Delta)}{\left(\widetilde{\Delta} - \Delta\right)^{1-\frac{c-c_0}{2}}}
\, e^{-2 \pi \, \widetilde{\Delta} \, \tau_2},
\end{equation}
where $\Gamma$ is the Euler gamma function.
This integral converges as long as $c>c_0$, which is the reason why we needed to impose $c>c_0$ for our deformation strategy to work.

Computing the truncated integral in \Eq{deformed_trunc} is now straightforward, though the intermediate results are not particularly enlightening.
The final result for the truncated, deformed partition function is:
\begin{equation}
\label{eq:deformed_truncated}
 Z_{\rm deformed}^{\rm trunc}(\tau; c; \Delta_{\rm max}) = \tau_2^{\frac{c_0 - c}{2}} \, \frac{|q|^{\frac{1-c}{12}}}{|\eta(\tau)|^2} \sum_{j,\Delta} \widetilde{a}_{\frac{\Delta+j}{2},\frac{\Delta-j}{2}} \, q^{\frac{\Delta+j}{2}} \,  \bar{q}^{\frac{\Delta-j}{2}} \, f(\tau_2,\Delta; \Delta_{\rm max}).
\end{equation}
Here, we have defined a form factor:
\begin{equation}
f(\tau_2, \Delta; \Delta_{\rm max}) =  \frac{\gamma\left(\tfrac{c-c_0}{2},2\pi(\Delta_{\rm max}-\Delta)\tau_2\right)}{\Gamma(\tfrac{c-c_0}{2})} \, \Theta(\Delta_{\rm max} - \Delta),
\end{equation}
where $\gamma$ is the lower incomplete Euler gamma function.
As expected, this form factor is zero if a term in the exact CFT partition function has $\Delta > \Delta_{\rm max}$.
When $\Delta_{\rm max}-\Delta$ is large compared to $1/\tau_2$, the form factor goes to 1 and we recover the untruncated deformed partition function.

\subsection{Estimating Truncation Uncertainties}

Putting these pieces together, we can estimate the truncation uncertainty $\sigma_{\rm trunc}$ in \Eq{improved_loss}.
This uncertainty depends sensitively on $\tau$, and it also depends on the central charge $c$ of the candidate CFT and on the truncation level $\Delta_{\rm max}$.
It also depends on the choice of baseline CFT, though we do not expect this choice to have too much of an impact, since all CFTs of the same central charge have the same Cardy estimate of the density of states.
This expectation will be borne out in \Tab{n1mm_m_vs_r_table} below.

Because there can be accidental cancellations between $Z^{\rm trunc}_{\rm deformed}(\tau)$ and $Z^{\rm trunc}_{\rm deformed}(\tau_S)$, we treat these terms as having independent truncation uncertainties, which we add in quadrature.%
\footnote{These accidental cancellations occur because sometimes $Z^{\rm trunc}_{\rm deformed}(\tau)$ is greater than $Z^{\rm trunc}_{\rm deformed}(\tau_S)$, and sometimes less, so there must be values of $\tau$ where the difference is zero.}
In addition, these independent truncation uncertainties can be accidentally small because of the oscillatory structure of the $j > 0$ characters, which we can mitigate by considering just imaginary values of $\tau$:
\begin{equation}
\label{eq:imaginary_mapping}
    \tau \to i \,|\tau|, \qquad \tau_S \to i \,|\tau_S| =  \frac{i}{|\tau|}.
\end{equation}
The final truncation uncertainty is therefore:
\begin{align}
\label{eq:final_trunc_unc}
\sigma_{\rm trunc}(\tau; c; \Delta_{\rm max})^2 &= \left(Z_{\rm deformed}(i\,|\tau|; c) - Z^{\rm trunc}_{\rm deformed}(i \,|\tau|; c; \Delta_{\rm max}) \right)^2 + (\tau \rightarrow \tau_S).
\end{align}
Recall that $Z_{\rm deformed}(\tau; c) = Z_{\rm deformed}(\tau_S; c)$ by construction, and the mapping in \Eq{imaginary_mapping} ensures that $Z_{\rm deformed}(i\,|\tau|; c)$ equals $Z_{\rm deformed}(i\,|\tau_S|; c)$.

While tedious, the steps to compute \Eq{final_trunc_unc} are easily automated in \mathematica.
To obtain $Z_{\rm deformed}^{\rm trunc}$, we need to determine the $\widetilde{a}_{h, \bar{h}}$ coefficients in \Eq{deformed_via_coeffs}, which requires not only extracting the ${a}_{h, \bar{h}}$ terms from the (estimate of the) partition function in \Eq{exact_k_kbar_expansion} but also doing a further expansion of $|\eta(\tau)|^2 \, |q|^{-\frac{1}{12}}$.
Because of a peculiarity in the way \mathematica\ performs power series expansions, the current version of our code can only handle integer values of $\Delta_{\rm max}$, though we expect this limitation to be overcome in future versions.

Finally, it is worth reemphasizing that $Z_{\rm deformed}$ does not correspond to an actual CFT.
Because of the logarithmic behavior of \Eq{deformed_cusp_limit}, this theory has a pathological vacuum, which can be seen by expanding \Eq{deformed_truncated} in the small $q$ limit:
\begin{equation}
     Z_{\rm deformed}^{\rm trunc}(\tau; c; \Delta_{\rm max}) =|q|^{-\frac{c}{12}} \left( \frac{\left(2 \pi \Delta_{\rm max} \right)^{\frac{c - c_0}{2}}}{\tfrac{c - c_0}{2} \Gamma\big(\tfrac{c - c_0}{2}\big)} + \ldots \right).
\end{equation}
This leading term corresponds to having a $\Delta_{\rm max}$-dependent vacuum degeneracy.
Luckily, this pathological vacuum behavior drops out of the uncertainty estimate in \Eq{final_trunc_unc}, which is why we think our approach is still a valid way to estimate the impact of truncation.
See \App{alpha_truncation} for an alternative deformation strategy with a completely valid vacuum state. 
From the ML perspective, it is important to emphasize that this uncertainty has been determined from theory alone and does not correspond to a learned quantity.

\section{Validation with $N=1$ Minimal Model}
\label{sec:n1mm}

To validate our approach to finding CFT candidates, we use five known CFTs to benchmark what CFT-like behavior looks like.
We focus on the $N=1$ minimal models (N1MMs) shown in \Fig{n1mm_spectrum} with $m \in \{5,6,7,8,9\}$, for which we can efficiently generate the CFT spectrum and corresponding torus partition function.
For most of this section, we focus on the $m = 5$ theory, whose central charge of $c = 81/70$ is the smallest from the N1MM family that is strictly greater than 1.
We pretend that the N1MM spectrum is the result of performing our ML-style procedure, which allows us to test every aspect of our analysis pipeline apart from the optimization itself.

\subsection{Order One Loss Implies CFT-like Behavior}

\begin{table}
\centering
\begin{tabular}{c | c  c c c c}
 & $m=5$ & $m=6$ & $m=7$ & $m=8$ & $m=9$ \\
\hline$R=1$ & $0.851$ & $1.16$ & $1.29$ & $1.40$ & $1.65$ \\
$\boldsymbol{R=\sqrt{2}}$ & $\mathbf{0.649}$ & $\mathbf{0.904}$ & $\mathbf{1.02}$ & $\mathbf{1.12}$ & $\mathbf{1.32}$ \\
$R=\sqrt{3}$ & $1.58$ & $1.89$ & $2.03$ & $2.13$ & $2.48$ \\
$R=2$ & $0.777$ & $1.03$ & $1.14$ & $1.24$ & $1.46$
\end{tabular}

\caption{
Loss values for the N1MM CFTs with $m \in \{5,6,7,8,9\}$, with truncation uncertainties estimated from the FBCB models with radii $R \in \{1,\sqrt{2},\sqrt{3},2\}$.
Here, we take $\Delta^{\rm N1MM}_{\rm max} = 4$, $\Delta^{\rm FBCB}_{\rm max} = 6$, and $\tau_2^{\rm max} = 6$ as our defaults.
All choices of $R$ shown yield stable $\mathcal{O}(1)$ loss values, but for concreteness we use $R = \sqrt{2}$ from \Fig{fbcb_spectrum_rsprt2} when estimating truncation uncertainties.
}
\label{tab:n1mm_m_vs_r_table}
\end{table}

The first validation we perform is to show that, once the loss is normalized by the truncation uncertainty as in \Eq{improved_loss}, a loss value of $\mathcal{O}(1)$ corresponds to CFT-like behavior.
In \Tab{n1mm_m_vs_r_table}, we show the value of the loss for the five N1MM benchmark models, varying which FBCB theory is used to estimate the truncation uncertainties.
As desired, these loss values are all around $1$, with factor of 2 variations among different choices for $m$ and the FBCB radius $R$.
We therefore conclude that our uncertainty estimate is reliable, at least for theories that are similar enough in spectrum to the N1MM models.
For concreteness, we take the default FBCB radius to be:
\begin{equation}
R = \sqrt{2},
\end{equation}
though through some internal tests, we found that other choices give similar results.
For the alternative uncertainty estimate in \App{alpha_truncation}, we find qualitatively similar (though generically a factor of 2-3 smaller) loss values.

\subsection{Loss Terms Follow Chi-Squared Distribution}
\label{sec:loss_follows_chi_squared}

\begin{figure}
    \centering
    \subfloat[][]{
        \label{fig:n1mm_loss_histo_naive}
        \includegraphics[width=0.45\linewidth]{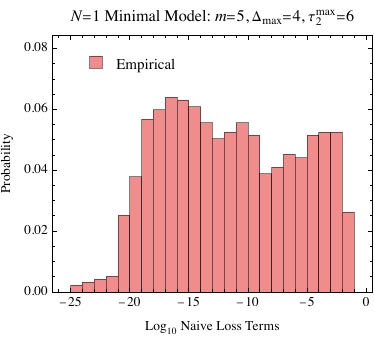}
    }
    $\qquad$
    \subfloat[][]{
        \label{fig:n1mm_loss_histo_improved}
        \includegraphics[width=0.45\linewidth]{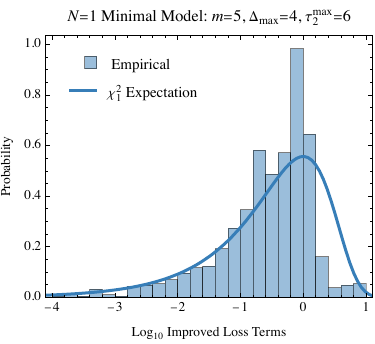}
    }
    \caption{Distribution of individual terms in the loss for (a) the naive MSE-style loss in red and (b) the improved $\chi^2$-style loss in blue, for the default N1MM with $m=5$. 
    For the naive loss, different values of $\tau$ yield vastly different loss terms, spanning 20 orders of magnitude in this case.
    For the improved loss, most values of $\tau$ yield order one loss values, with the small-side tails reasonably well modeled by a chi-squared distribution with 1 degree of freedom.
    The high-side depletion arises because the loss values are somewhat correlated, so that when one value of $\tau$ happens to yield a small loss, there are multiple nearby $\tau$ values that also have small losses.
    }
    \label{fig:n1mm_loss_histo}
\end{figure}

As discussed in \Sec{loss}, one of the motivations for introducing the uncertainty-improved loss was to avoid excessive sensitivity to the choice of sampled $\tau$ values.
In \Fig{n1mm_loss_histo_naive}, we show the individual terms that go into the naive loss in \Eq{naive_loss} for the default N1MM scenario.
We see that they span 20 orders of magnitude, implying that they are highly sensitive to the particular values of $\tau$ being sampled.

In \Fig{n1mm_loss_histo_improved}, we show the individual terms that enter the improved loss from \Eq{improved_loss}.
The individual terms are now highly concentrated around $\mathcal{O}(1)$ values.
If we were performing a $\chi^2$ analysis in a setting where the uncertainties were entirely statistical in nature, we would expect the individual terms in the loss to follow a chi-squared distribution with one degree of freedom.
The case of truncating a torus partition function is far from statistical; nevertheless, this chi-squared expectation for the improved loss is borne out in \Fig{n1mm_loss_histo_improved}.
While there is some depletion from the high-side tail, the low-side tail follows the expectation from a chi-squared distribution.

Because of this strong agreement with the chi-squared distribution, we will interpret the loss in a quasi-statistical framework.
We already did so in \Sec{parameter_uncertainties}, where we derived parameter uncertainties according to the second derivative of the loss.
We will also do so when defining an absolute measure of CFT-ness in \Eq{confidence_thresholds} below, where we define  ``$n\sigma$'' thresholds on the loss of:
\begin{equation}
    L < n^2,
    \label{eq:n_sigma}
\end{equation}
according to how typical the loss value is under a chi-squared distribution with 1 degree of freedom.
The number of degrees of freedom should not be taken literally, since the individual terms in the loss are highly correlated, and when one term in the loss is high/low, there are often nearby $\tau$ values exhibiting the same behavior.
Similarly, the degree of confidence is really a subjective criterion, coming from the human experience of doing many optimization runs and seeing what configurations are stable.
We have some evidence that the loss values tend to follow a chi-squared distribution with 3--5 degrees of freedom, albeit with an inflated high-side tail, but we nevertheless stick with \Eq{n_sigma} for reporting the results, keeping in mind that these are somewhat arbitrary thresholds.

\subsection{Stability to Truncation Choices}

\begin{figure}
    \centering
    \subfloat[][]{
        \label{fig:n1mm_cut_scan_naive}
        \includegraphics[width=0.45\linewidth]{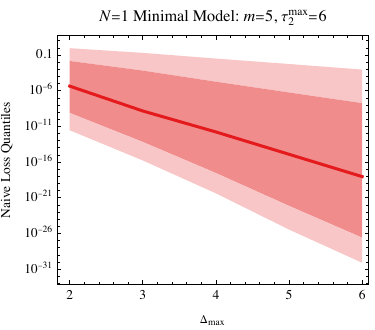}
    }
    $\qquad$
    \subfloat[][]{
        \label{fig:n1mm_cut_scan_improved}
        \includegraphics[width=0.45\linewidth]{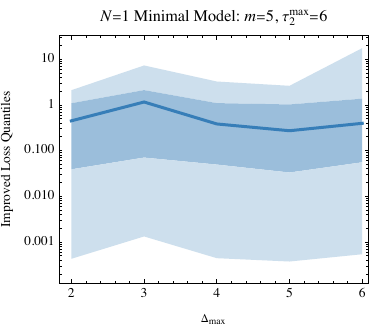}
    }

    \subfloat[][]{
        \label{fig:n1mm_tau2_scan_naive}
        \includegraphics[width=0.45\linewidth]{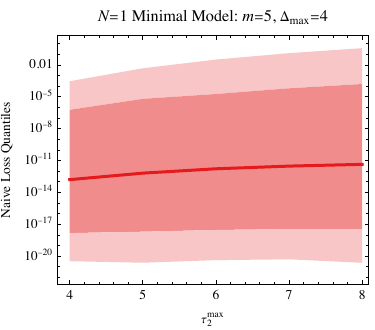}
    }
    $\qquad$
    \subfloat[][]{
        \label{fig:n1mm_tau2_scan_improved}
        \includegraphics[width=0.45\linewidth]{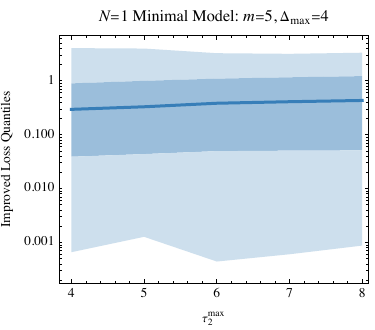}
    }

    \caption{
    Impact of different truncation choices on the distribution of loss terms, with the median loss shown as the solid line, the central 68\% interval as the darker shading, and the central 95\% interval as the lighter shading, for the default N1MM with $m=5$.
    The top plots scan the truncation in operator dimension for $\Delta_{\rm max} \in \{2,3,\mathbf{4},5,6\}$,
    while the bottom plots scan the truncation in the fundamental domain for $\tau_{2}^{\rm max} \in \{4,5,\mathbf{6},7,8\}$, with the bold number indicating the choice made for the complementary scan.
    When estimating the truncation uncertainties, we always set $\Delta^{\rm FBCB}_{\rm max} = \Delta_{\rm max} + 2$.
    For the naive loss in the left (red) plots, there is a high degree of sensitivity to the truncation, with an exponential decrease in the median loss with increasing $\Delta_{\rm max}$, and a nearly exponential increase in the upper edge of the 95\% interval with increasing $\tau_{2}^{\rm max}$.
    For the improved loss in the right (blue) plots, the distributions are much more stable across truncation choices, with less than an order of magnitude change in loss quantiles among the variations.
    The sharpest change in the improved loss is between $\Delta_{\rm max} = 3$ and $\Delta_{\rm max} = 4$, due to the presence of multiple primary operators with $\Delta \approx 3$ in the default N1MM, as can be seen in \Fig{n1mm_spectrum_m5}.
}
    \label{fig:n1mm_truncation_scans}
\end{figure}

Another motivation for introducing the uncertainty-improved loss was to mitigate sensitivity to the truncation scales $\Delta_{\rm max}$ and $\tau_2^{\rm max}$.
In the left-hand panels of \Fig{n1mm_truncation_scans}, we show the impact of truncation using the naive loss, evaluated on the default N1MM model.
Here, we are showing quantiles of the individual loss terms, so we can track both the median loss values and their spread.
As expected, when the $\Delta_{\rm max}$ scale increases, the naive loss decreases exponentially, since the spectrum is becoming more and more CFT-like.
Nevertheless, the variance of the loss values increases with increasing $\Delta_{\rm max}$, spanning tens of orders of magnitude, highlighting why the naive loss is not a reliable measure of CFT-ness.
The sensitivity to $\tau_2^{\rm max}$ is less dramatic, but one still sees a noticeable increase in extreme loss values as $\tau_2^{\rm max}$ increases.

With the improved loss, shown in the right-hand panels of \Fig{n1mm_truncation_scans}, the stability with truncation is much improved.
First, notice the difference in scale on the $y$-axis, where the improved loss quantiles are contained within 4 orders of magnitude.
As $\Delta_{\rm max}$ is varied, the loss does change in an apparent way, but this reflects actual features in the spectrum, in particular the presence of multiple primary operators with $\Delta \approx 3$.
Apart from some low-side variability, there is barely any impact from changing $\tau_2^{\rm max}$.
Thus, we conclude that the improved loss is more resilient to the way we cut off of the spectrum and sampling ranges.

\subsection{Highly Correlated Parameter Uncertainties}
\label{sec:n1mm_para_uncertainties}

As our final validation of our methodology, we estimate the uncertainties on the parameters of the default N1MM model.
The truncated N1MM model with $m=5$ does \emph{not} correspond to a local minimum of the loss landscape.
Still, because we are using the approximate inverse covariance matrix in \Eq{covariance_approx} that is positive semi-definite by construction, we can derive uncertainties on the parameters that faithfully quantify how far we can move in parameter space and still achieve an order 1 value for the loss.
Here, we vary both the central charge $c$ and operator dimensions $\{\Delta_a\}$, but not the discrete parameters.

The truncated N1MM model with $m=5$ involves 41 parameters.%
\footnote{There is a current operator with $\Delta = |j| = 4$, but this is removed by the $\Delta_{\rm max} = 4$ constraint, which is based on strict inequality.  If we included this operator, we could not assess an uncertainty on its dimension, because of the discontinuous behavior of the Virasoro characters from \Eq{virasoro_characters} at $h = 0$ and $\bar{h} = 0$.}
Taking into account numerical accuracy, though, the rank of the inverse covariance matrix is only around 10, implying that there are around 30 flat directions in the loss landscape.
This is very strong evidence for a continuous space of solutions.%
\footnote{One might worry that this continuous space of solutions is just an artifact of working with the truncated bootstrap equations.
While we cannot exclude the possibility that the true untruncated solutions to the modular bootstrap equations are isolated, we do see evidence that as the truncation scale increases, the number of flat directions also increases.
For example, raising the truncation scale to $\Delta_{\rm max} = 5$, the rank of the inverse covariance matrix stays around 10, even though the number of free parameters increases to 59.}
For the non-zero eigenvalues, they span around 15 orders of magnitude, so the ``canyon walls'' around the loss valley have an exponential hierarchy of scales.
Because of this hierarchical structure, the covariance matrix itself is not particularly illuminating.

\begin{figure}[p]
    \centering
    \includegraphics[width=0.7\linewidth]{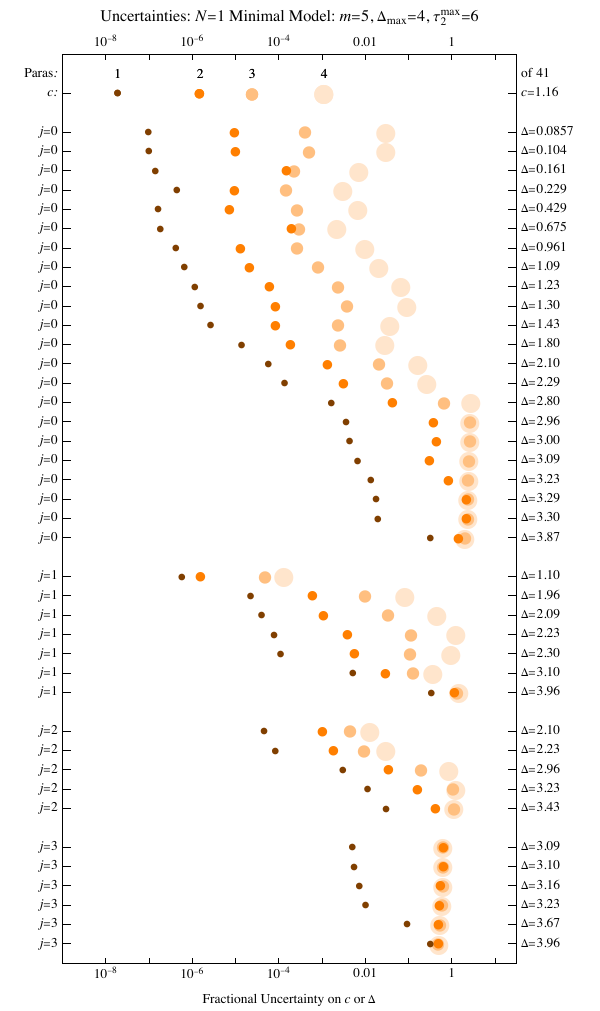}
    \caption{Fractional uncertainties on the central charges and primary dimensions for the default N1MM model with $m=5$ from \Fig{n1mm_spectrum_m5}, as estimated from the derivatives of the loss function.
    The uncertainties are highly correlated, so we show the maximum excursion on each parameter as 1, 2, 3, and 4 of them are varied simultaneously.
    When only 1 parameter is varied, the central charge and lowest dimension operators are constrained to better than 1 part in a million, but this neglects important correlations between the parameters. 
    }
    \label{fig:n1mm_uncertainty}
\end{figure}

As a more intuitive way to understand the hierarchical uncertainty structure, we show the impact of varying a small number of parameters in \Fig{n1mm_uncertainty}.
To make this plot, we constructed submatrices of the inverse covariance matrix $\Sigma^{-1}$, took the inverse to determine the partial covariance matrices $\{\Sigma_{\rm sub}\}$, and then found the maximum diagonal partial covariance for each parameter.
When only 1 parameter is varied, the uncertainty on parameter $a$ is simply:
\begin{equation}
    \sigma_a = \sqrt{\Sigma^{\rm sub}_{aa}} = \frac{1}{\sqrt{(\Sigma^{-1})_{aa}}}.
\end{equation}
For the central charge and lowest lying operators, we see constraints that are better than a part per million, indicating a highly narrow loss canyon.
As more parameters are varied simultaneously, though, the constraints on individual operators degrade significantly, with constraints loosening to the few percent level after four operators are varied.
This shows that the space of solutions exhibits highly complicated correlations among the parameters.

\section{Hunting for New CFTs}
\label{sec:hunt}

We now embark on a hunt for candidate CFT partition functions, exploring the $c \in (1,\frac{8}{7})$ region with no known CFT examples.
This study will provide strong evidence for a continuous space of solutions to the modular bootstrap equations, motivating the more extensive survey in \Sec{survey}.

\subsection{Hunting Strategy}
\label{sec:hunting_strategy}

To try to identify the highest quality CFT candidates we can, we want to impose as few constraints on the parameters as possible.
As explained in \Sec{partition_function}, it makes sense to fix the central charge $c$, since this has a large impact on the truncation uncertainty estimate.
Concretely, we fix five benchmark central charge values for this hunt:
\begin{equation}
c \in \{1.05,1.07,1.09,1.11,1.13\},
\end{equation}
though it is straightforward to run the optimization for any value of $c$.

\begin{figure}
    \centering
    \includegraphics[width=0.5\linewidth]{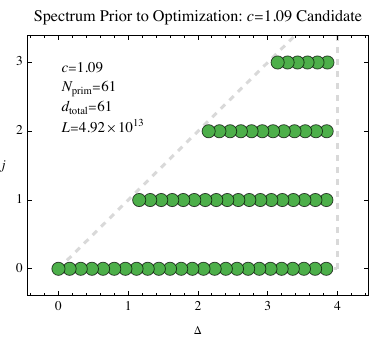}
    \caption{Initial spectrum of primary operators prior to optimization.
    This model has 60 free parameters, namely the dimensions $\{\Delta_a\}$ of its 60 non-vacuum primaries.
    Note that the central charge is \emph{not} a free parameter during the optimization, and while the $c = 1.09$ initialization is shown here for concreteness, the same initial spectrum is used for all $c$ values, though with different starting loss $L$.
    }
    \label{fig:hunt_starting_plot}
\end{figure}

Other than this central charge constraint, we impose the bare minimum of restrictions.
We fix the truncation scale at $\Delta_{\rm max} = 4$, and initialize the optimization with 60 operators with spins $j \in \{0,1,2,3\}$:
\begin{equation}
\label{eq:hunt_num_operators}
    N_{j=0} = 24, \qquad     N_{j=1} = 18, \qquad     N_{j=2} = 12, \qquad     N_{j=3} = 6.
\end{equation}
This is typically more operators than needed to find modular-invariant partition functions in this $c$ range, but we let any extraneous operators drift above the $\Delta_{\rm max}$ threshold during the optimization as needed.

The initial values of the operator dimensions $\{\Delta_a\}$ are shown in \Fig{hunt_starting_plot}, which are common for all values of $c$.
We do not impose any restrictions on $\Delta_a$ apart from $\Delta_a \ge |j_a|$, as required by unitarity.
We then optimize the operator dimensions according to the procedure in \Sec{ml}, focusing on the \sven\ algorithm with default parameters from \Eq{sven_default}.
We perform 300 iterations of training, which is well after all of the optimization algorithms have converged.

\subsection{Evolution of Loss Function}

\begin{figure}
    \centering
    \subfloat[][]{
        \includegraphics[width=0.45\linewidth]{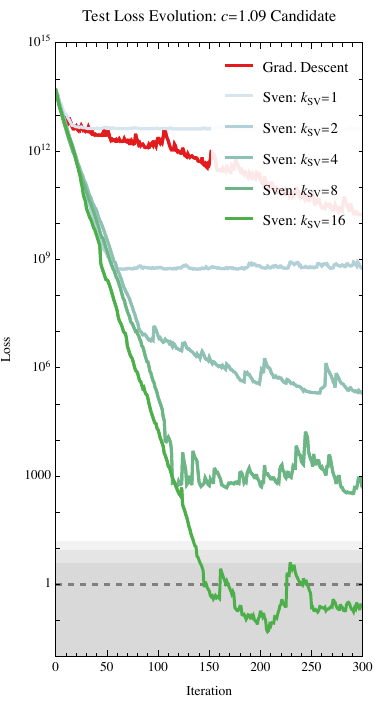}
        \label{fig:campaign_loss_vs_iter_svgrad}  
    }$\qquad$
    \subfloat[][]{
        \includegraphics[width=0.45\linewidth]{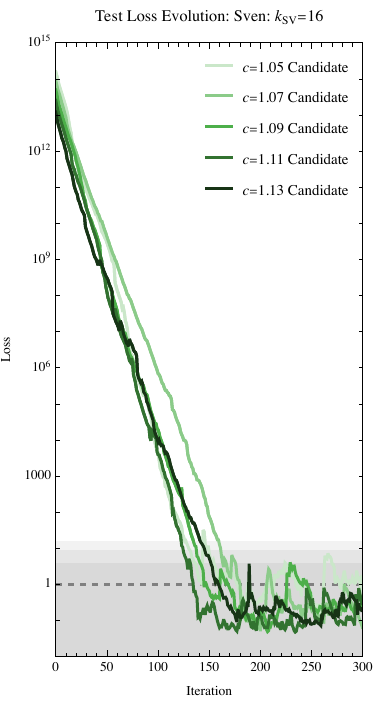}
        \label{fig:campaign_loss_vs_iter_all}  
    } 
    \caption{
    Evolution of the test loss during training, with the gray bars at the bottom indicating the CFT-like thresholds from \Eq{n_sigma} with $n \in \{2,3,4\}$. 
    In the left panel, we compare different optimization strategies for the $c=1.09$ candidate.
    Whereas gradient descent only yields a modest decline in the loss, \sven\ is able to find orders of magnitude deeper minima as the number of computed singular values $k_{\rm SV}$ increases.
    For the default value of $k_{\rm SV} = 16$, the test loss crosses the CFT-like threshold in around 150 steps.
    In the right panel, we see that the training dynamics for the five candidate theories exhibit qualitatively similar behaviors.
    Note that we are only showing the test loss, but as explained in the text, the training loss faithfully tracks the test loss up to statistical fluctuations.}
    \label{fig:campaign_loss_vs_iter}  
\end{figure}

During training, we evaluate the test loss to assess how well the optimization is performing.
Because we are using stochastic sampling of the $\tau$ values during training, the test loss and training loss track each other closely, with the observed differences consistent with statistical fluctuations.
No aspect of the training depends on the test set, so the test loss offers an unbiased estimate of the true loss (albeit without the $\Delta_{\rm max}$ constraint).
To get a more accurate estimate of the final loss at the end of training (including the $\Delta_{\rm max}$ constraint), we use a separate reporting sample for \Tab{campaign_m_vs_c_table} below.

In \Fig{campaign_loss_vs_iter_svgrad}, we compare the test loss evolution for different optimization algorithms, all with the same $\eta = 0.1$ learning rate, focusing on the $c = 1.09$ candidate for concreteness.
Gradient descent has a difficult time making any optimization progress, and even if we run it with 10 times more iterations, it plateaus well above the CFT-like target.
\sven\ with just $k_\text{SV} = 1$ singular value fares even worse, asymptoting to a bad solution after tens of steps.
As the number of singular values is increased, though, \sven\ is able to find deeper and deeper minima, ultimately finding a high-quality CFT candidate using $k_\text{SV} = 16$.
As shown in \Fig{campaign_loss_vs_iter_all}, this $k_\text{SV} = 16$ default value works for all five $c$ values we considered, with a good solution found in roughly 150 iterations.

Even after crossing the CFT-like threshold, the optimization exhibits stochastic loss fluctuations.
These fluctuations are roughly the same size as the loss thresholds in \Eq{confidence_thresholds}, providing further evidence that order 1 losses are a robust indication of CFT-like behavior.
We checked that using $k_\text{SV} = 32$ does not result in a better minimum, though this is partly due to the fact that we only use $N_\tau^{\rm train} = 32$ samples in each training iteration, so the additional singular values are very noisy.
We also tried adjusting the learning rate, finding that $\eta = 0.03$ slows the learning but does not qualitatively change the solutions, whereas $\eta = 0.3$ often causes operators to drift past the $\Delta_{\rm max}$ boundary too quickly.

\begin{table}
\centering
\begin{tabular}{c | c  c c c c}
 & $c=1.05$ & $c=1.07$ & $c=1.09$ & $c=1.11$ & $c=1.13$ \\
\hline$R=1$ & $1.19$ & $0.528$ & $0.680$ & $3.02$ & $2.39$ \\
$\boldsymbol{R=\sqrt{2}}$ & $\mathbf{0.878}$ & $\mathbf{0.395}$ & $\mathbf{0.510}$ & $\mathbf{2.24}$ & $\mathbf{1.80}$ \\
$R=\sqrt{3}$ & $3.39$ & $1.30$ & $1.61$ & $6.64$ & $5.11$ \\
$R=2$ & $1.14$ & $0.513$ & $0.649$ & $2.78$ & $2.18$
\end{tabular}

\caption{Same as \Tab{n1mm_m_vs_r_table}, but now for the five candidate CFTs found through optimization.
We see the same stability as in the N1MM study as a function of $R$ and the choice of model, though some of the loss values are a factor of 2 smaller or larger.
While this is some evidence for overfitting (for smaller losses) and incomplete optimization (for larger losses), we do not expect this to impact the qualitative features of the identified spectra.
}
\label{tab:campaign_m_vs_c_table}
\end{table}

In \Tab{campaign_m_vs_c_table}, we show the final reported values of the loss, using different baseline FBCB models to estimate truncation uncertainties.
These loss values are similar to the ones found for the N1MM benchmarks from \Tab{n1mm_m_vs_r_table}, giving us confidence that we have indeed found CFT-like solutions.
Some of the loss values tend to be a factor of 2 smaller or larger than the N1MM benchmarks, which is some evidence for overfitting or incomplete optimization, respectively.
Nevertheless, because of how sensitive the loss is to small changes in the spectrum, we do not think that these are causing significant distortions in the found spectra.

As mentioned in \Sec{sven}, we use an adaptive method to choose the number of singular value directions \sven\ explores in each iteration.
At the beginning of training, \sven\ tends to use most of the $k_{\rm SV} =16$ singular values, tapering down to roughly half that at the end of training.
Apparently, the ability of \sven\ to explore multiple parameter directions early in the training is important for finding good CFT candidates, and the primary role of adaptation is to make sure that noisier directions are ignored later in the training.

\subsection{Candidate CFT Partition Functions}

\begin{figure}
    \centering
    \subfloat[][]{
        \includegraphics[width=0.4\linewidth]{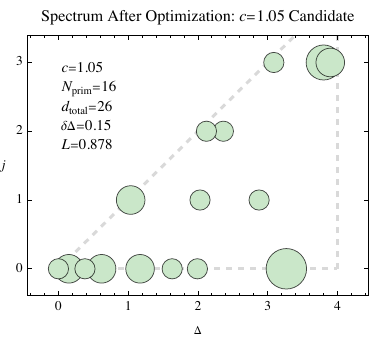}
    }$\qquad$
    \subfloat[][]{
        \includegraphics[width=0.4\linewidth]{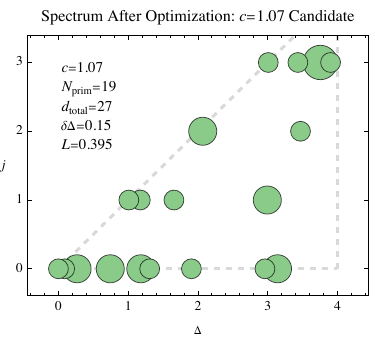}
    }\\
    \subfloat[][]{
        \includegraphics[width=0.4\linewidth]{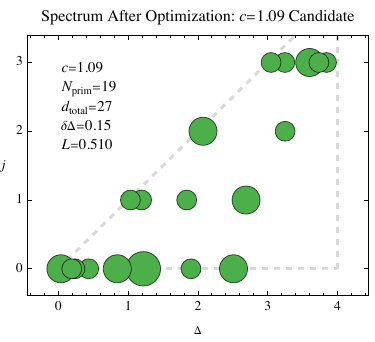}
        \label{fig:campaign_ending_plot_109}
    }\\
    \subfloat[][]{
        \includegraphics[width=0.4\linewidth]{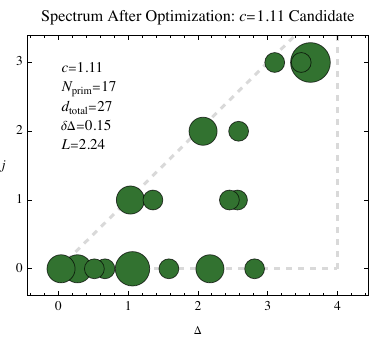}
    }$\qquad$
    \subfloat[][]{
        \includegraphics[width=0.4\linewidth]{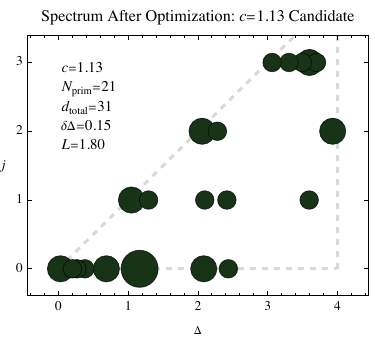}
    }
    \caption{Final spectrum of Virasoro primary operators after optimization, where we have merged operators with $|\Delta_a - \Delta_b| < 0.15$ for ease of visualization.
    The fact that these spectra are qualitatively similar to the N1MM models in \Fig{n1mm_spectrum} is additional evidence for their CFT-like structure.
    The presence of non-vacuum primaries with $\Delta \approx |j|$ suggests that these candidates could be in the vicinity of rational CFTs.}
    \label{fig:campaign_ending_plot}    
\end{figure}

After this evidence for successful optimization, we can now inspect the resulting candidate CFT spectra, as shown in \Fig{campaign_ending_plot}.
To make it easier to read these plots, we have combined operators whose dimensions are within $|\Delta_a - \Delta_b| < 0.15$ of each other, to build an effective operator with degeneracy greater than 1.
We emphasize, though, that the degeneracies are fixed at 1 throughout the training.
Because the optimization lets operators drift past the $\Delta_{\rm max}$ boundary, the original 60 operators have been cut down by roughly half.

Visually, these candidates look roughly as CFT-like as the N1MM scenarios from \Fig{n1mm_spectrum}, with clusters of primaries separated by noticeable gaps in the spectrum.
There is perhaps a depletion of operators with $\Delta \approx 3.5$ compared to expectation, which is most noticeable in the $c= 1.09$ and $c=1.11$ models.
That said, operators close to the $\Delta_{\rm max} = 4$ boundary are poorly constrained, so it is possible that some ``true'' primaries accidentally drifted above the $\Delta_{\rm max}$ threshold.
Cardy scaling suggests that the number of primary operators should increase with $c$, and there is mild evidence for this trend among these five examples.
For completeness, we show the spectra for the incomplete optimizations from \Fig{campaign_loss_vs_iter_svgrad} in \App{additional_visualization}.

\begin{figure}[p]
    \centering
        \includegraphics[width=0.7\linewidth]{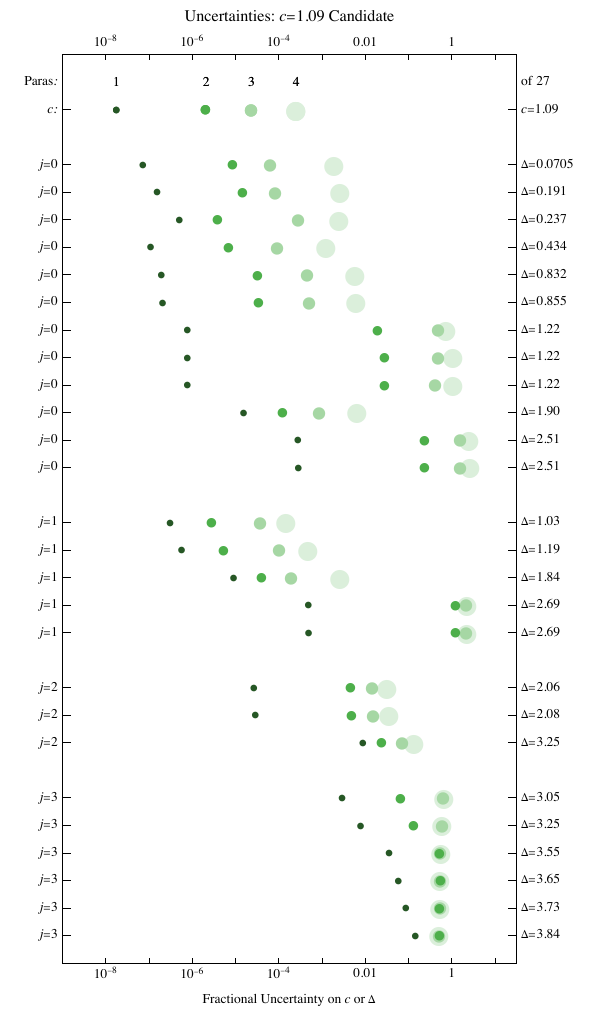}
    \caption{
Fractional uncertainties on the central charges and primary dimensions for the $c = 1.09$ candidate, as estimated from the derivatives of the loss function.
While the central charge was fixed during training, we allow it to float when reporting uncertainties.
Even if there is evidence for operator degeneracy, we vary their dimensions independently.
Like for the N1MM benchmark in \Fig{n1mm_uncertainty}, the central charge and lowest primary dimensions have less than 1 part per million uncertainties when varied individually, but there are sizable correlated uncertainties when multiple parameters are varied simultaneously.
The uncertainties for the other four benchmark models (see \Figs{campaign_uncertainty_105_107}{campaign_uncertainty_111_113}) are qualitatively similar.
    }
    \label{fig:campaign_uncertainty_109}    
\end{figure}

We can inspect the robustness of these solutions by studying the parameter uncertainties.
We focus on the $c = 1.09$ candidate in \Fig{campaign_uncertainty_109}, presenting the other candidates in \App{additional_visualization}.
Just as for the N1MM benchmark in \Fig{n1mm_uncertainty}, the central charge and lowest dimension primaries are highly constrained, with better than part per million accuracy.
As more operator dimensions are varied simultaneously, though, the uncertainties grow, providing further evidence for a continuous space of solutions to the modular bootstrap equations.
As expected, the higher-spin operators are rather poorly constrained, with uncertainties large enough in some cases that we cannot confidently claim that they are below $\Delta_{\rm max}$.
Correspondingly, there may be operators that drifted above the $\Delta_{\rm max}$ boundary during training whose actual dimension is below the $\Delta_{\rm max}$ threshold.
To avoid accidentally small uncertainties, we include operators with $\Delta_a > \Delta_{\rm max}$ when constructing $\Sigma^{-1}$, though we do not report their uncertainties (which are always order 1).

An interesting feature of these spectra is that they all have at least one operator with $\Delta_a \approx |j_a|$.
This is not surprising from the optimization perspective, since even if there is no physical reason to prefer $\Delta_a = |j_a|$, the optimization can naturally end up there, since it is at the boundary of the allowed parameter space.
None of the $\Delta_a \approx |j_a|$ operators are as close to the boundary as their single parameter uncertainties, but many of them are consistent with $\Delta_a = |j_a|$ when multiple parameters are varied.

We regard this hunting campaign as successful, in the sense of finding high-quality CFT candidates.
Still, these results are not particularly informative about the generic properties of  CFTs at small central charge.
At this point it seems to be straightforward to find solutions to the modular bootstrap equations with $c\in(1,\frac{8}{7})$.
To draw stronger conclusions, we need to impose additional constraints on the spectrum to see if we can find the edges of the allowed parameter space.

\section{Surveying the Space of Low $c$ CFTs}
\label{sec:survey}

Having established the power of our ML-style optimization to find candidate CFTs, we now explore a broader space of possible spectra.
In this section, we survey the space of central charge versus spectral gap:
\begin{equation}
    (c,\Delta_{\rm gap}),
\end{equation}
looking to see if any interesting relationships emerge between $c$ and $\Delta_{\rm gap}$.
At minimum, we expect our candidate CFTs to satisfy the HCLY bound in \Eq{hellerman}.
As already previewed in \Fig{summary_plots}, we encounter a fascinating obstruction to finding CFT candidates with $c \in (1.00,1.06)$ and $\Delta_{\rm gap} \in (0.3,0.5)$, suggesting that it may be possible to derive a stronger dual constraint on $\Delta_{\rm gap}$.
Moreover, we find evidence that this constraint might grow even stronger with increasing $\Delta_{\rm max}$.

\subsection{Surveying Strategy}

For this survey, we follow a similar approach as in \Sec{hunting_strategy}, just with an additional constraint on the spectral gap.
We consider 10 equally-spaced values for the central charge:
\begin{equation}
    c \in \{1.02,1.04, \ldots, 1.18,1.20\},
\end{equation}
and 15 equally-spaced values for the spectral gap $\Delta_{\rm gap}$:
\begin{equation}
    \Delta_{\rm gap} \in \{0.04,0.08, \ldots, 0.56,0.60\},
\end{equation}
for a total of 150 configurations.
The only thing that really limits how finely we scan is computational time, since as we saw from the parameter uncertainty studies in \Figs{n1mm_uncertainty}{campaign_uncertainty_109}, $c$ and $\Delta_{\rm gap}$ can be individually constrained to less than a part per million.%
\footnote{In preliminary studies that scan the central charge very close to $c = 1$, we find that the bounds on $\Delta_{\rm gap}$ become less stringent.  This is consistent with the expectation that one should be able to find solutions to the truncated modular bootstrap equations by deforming a free boson theory to have $c = 1 +\epsilon$ along with corresponding $\mathcal{O}(\epsilon)$ changes to the operator dimensions, though this strategy requires $\epsilon \lesssim 10^{-2}$.}

\begin{figure}
    \centering
    \includegraphics[width=0.5\linewidth]{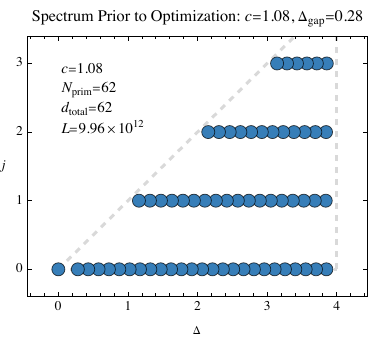}
    \caption{Same initial spectrum prior to optimization as \Fig{hunt_starting_plot}, but now imposing a non-trivial spectral gap $\Delta_{\rm gap}$.
    This model will evolve into the strong candidate CFT shown in \Fig{survey_yes_high_example}.
    }
    \label{fig:survey_starting_plot}
\end{figure}

To fix the spectral gap $\Delta_{\rm gap}$, we force the first non-vacuum primary to have:
\begin{equation}
    \label{eq:survey_fixed_operator}
    (\Delta_1, j_1,d_1) = (\Delta_{\rm gap},0,1).
\end{equation}
We then use the strategy in \Eq{map_to_gap} to constrain the rest of the primaries to have $\Delta_a > \max(|j_a|,\Delta_{\rm gap})$.
Note that there is no gap between $\Delta_1$ and the next highest primary, so the optimization can effectively increase the degeneracy of $\Delta_1$ by setting additional operator dimensions close to $\Delta_{\rm gap}$.
We initialize the optimization with the same spins and number of operators as in \Eq{hunt_num_operators}, just with the extra fixed operator from \Eq{survey_fixed_operator}, and again let operators drift above $\Delta_{\rm max}$ if they are not needed.
An example primary spectrum at initialization is shown in \Fig{survey_starting_plot}.

As one increases the number of constraints on the model, it becomes increasingly difficult to find high-quality solutions.
We therefore expect somewhat larger loss values for this survey than the study in \Sec{hunt} where $\Delta_{\rm gap}$ was unconstrained.
To partially counteract this, we use the testing $\tau$ sample to select the best test loss value through the whole training run, since often there is a ``better'' solution at an intermediate stage of the training.
To avoid selection bias, we use the reporting $\tau$ samples to compute the final loss, and use the loss thresholds in \Eq{confidence_thresholds} to report our ``confidence'' in the results.
Had we used the testing loss instead, we would have been overconfident in the solutions.

\subsection{Results of the Survey}
\label{sec:survey_results}

\begin{figure}
    \centering

    \subfloat[][]{
\includegraphics[width=0.45\linewidth]{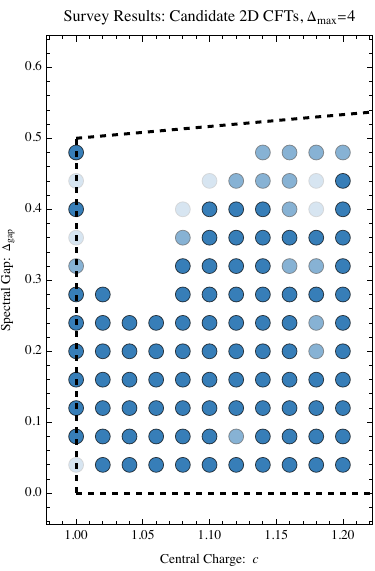}
\label{fig:survey_results_yes}
    } $\qquad$        
  \subfloat[][]{
\includegraphics[width=0.45\linewidth]{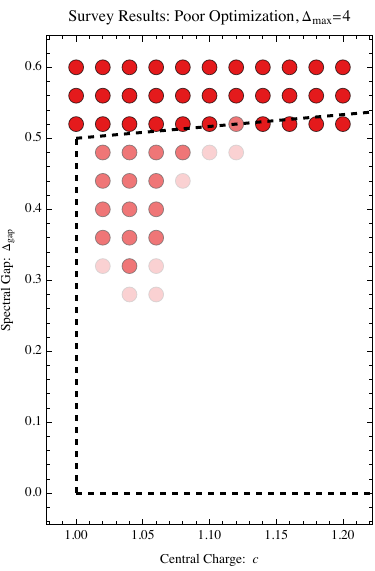}
\label{fig:survey_results_no}
    }
    \caption{Results of the $(c,\Delta_{\rm gap})$ survey in the plane of central charge versus spectral gap, separating the preview plot from \Fig{intro_yesno_examples}.
    In the left panel, we show candidate CFTs identified in our survey.
    The blue dots indicate the degree of confidence in the solution: darkest blue for loss $L < 4$, medium blue for $L \in (4,9)$, and the lightest blue for $L \in (9,16)$.
    In the right panel, we show poor optimization results indicative of a modular bootstrap obstruction.
    The red dots indicate the degree of deficiency in the optimization: 
darkest red for loss $L > 10^7$, medium red for $L \in (10^3,10^7)$, and  lightest red for $L \in (16,10^3)$.
    The dashed black lines correspond to the unitarity bound $\Delta_{\rm gap} \ge 0$ and HCLY bound $\Delta_{\rm gap} \le \frac{c}{6}+\frac{1}{3}$~\cite{Collier:2016cls}.
    }
    \label{fig:survey_results}
\end{figure}

We now show the results of our $(c,\Delta_{\rm gap})$ survey, separating out the preview plot from \Fig{intro_yesno_examples} into positive and negative examples.
Following \Eq{n_sigma}, we use the following criteria to define good quality CFT candidates:
\begin{align}
    \text{High confidence (``$<2 \sigma$'')}&: \quad L < 4,\nonumber\\
    \text{Medium confidence (``$<3 \sigma$'')}&: \quad L \in (4,9),\nonumber\\
    \text{Low confidence (``$<4 \sigma$'')}&: \quad L \in (9,16),
\label{eq:confidence_thresholds}
\end{align}
where ``$\sigma$'' should not be taken too literally, as discussed in \Sec{loss_follows_chi_squared}.
In \Fig{survey_results_yes}, we indicate in blue spectra that satisfy these criteria, with light/medium/dark blue corresponding to low/medium/high confidence.
These examples cover much of the surveyed parameter space, albeit with confidence levels that fluctuate due to variability in the training.
The continuity of the good solutions is further evidence for a space of solutions, and we expect that with further optimization, lower confidence solutions could turn into higher confidence ones.
Because the optimization problem is harder the closer the spectra are to the HCLY bound, it is understandable that the quality of the solution degrades with increasing $\Delta_{\rm gap}$.

In \Fig{survey_results_no}, we indicate in red spectra that do not satisfy the loss criteria in \Eq{confidence_thresholds} so are unlikely to be (nor be in the vicinity of) a candidate CFT.
Light/medium/dark red correspond to weak/intermediate/strong exclusion, with somewhat arbitrary thresholds of: 
\begin{align}
    \text{Weak exclusion}&: \quad L \in (16,10^3),\nonumber\\
    \text{Intermediate exclusion}&: \quad L \in (10^3,10^7),\nonumber\\
    \text{Strong exclusion}&: \quad L > 10^7.
\end{align}
As expected, all of the points above the HCLY bound from \Eq{hellerman} are strongly excluded.
Of course, because we are working in the primal bootstrap, the inability to find a solution does not prove that one does not exist, but it does provide strong circumstantial evidence.
At the boundary between the good and bad solutions, there are a handful of weakly-excluded solutions, so it is possible that further optimization could improve the result.
It is also possible that these weakly-excluded solutions (along with some of the low confidence examples) could become worse as $\Delta_{\rm max}$ is increased, which is something we will study in \Sec{deltamax5} below.

Most strikingly, the upper left corner of \Fig{survey_results_no} has multiple spectra that satisfy the HCLY bound but are very far from satisfying our primal modular bootstrap constraints.
These are at the intermediate exclusion level, suggesting that the spectrum has some features that are CFT-like, but modular invariance does not hold.
Again, it is possible that we simply need to use different optimization parameters or a better optimization algorithm to find examples in this region.
Still, this ``gap'' of non-solutions between the primal examples and the known dual bound is a key result from this study.
The apparent existence of a primal/dual gap suggests a fundamental obstruction from the modular bootstrap that constrains the space of possible CFTs.

\begin{figure}
    \centering
    \includegraphics[width=0.5\linewidth]{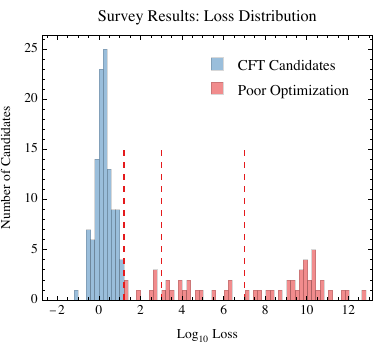}
    \caption{Distribution of loss values found in the $(c,\Delta_{\rm gap})$ survey.
    The losses for the CFT candidates are peaked near 1 (logarithm peaked near 0), as expected for chi-squared-like behavior, albeit with an asymmetric skew towards higher loss values.
    While the boundary between CFT candidates and weakly-excluded spectra is fluid, losses corresponding to intermediate/strong exclusion are well separated from those of good solutions.
    }
    \label{fig:survey_loss_distribution}
\end{figure}

To gain further confidence in the solutions, it is instructive to look at the distribution of loss values, shown in \Fig{survey_loss_distribution}.
We see that the CFT candidates have loss values that are strongly peaked near $L = 1$ (i.e.\ $\log_{10} L = 0$), with a skewed distribution to somewhat larger loss values.
This upward skew is expected because the training loss used to select the ``best'' parameters did not include the $\Delta_{\rm max}$ truncation, whereas the reporting loss does.
If we had used the testing loss instead of the reporting loss, then the loss values would have been shifted to lower values, with more values less than one compared to a chi-squared-like expectation.
This plot makes clear that the boundary between ``low confidence'' candidates and ``weakly excluded'' spectra is fluid, and it is hard to say whether we are seeing a high-side tail of good candidates or a low-side tail of bad ones.
On the other hand, the intermediate/strong exclusions are clearly separated from the good solutions, keeping in mind that this is plotted on a logarithmic scale.

\begin{figure}
    \centering
    \subfloat[][]{
    \includegraphics[width=0.4\linewidth]{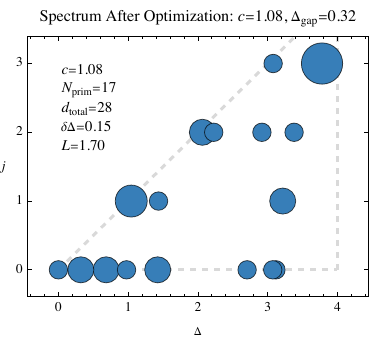}
    \label{fig:survey_yes_high_example}
    }$\qquad$
        \subfloat[][]{
    \includegraphics[width=0.4\linewidth]{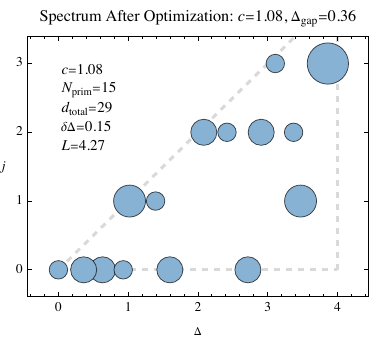}
    }\\
        \subfloat[][]{
    \includegraphics[width=0.4\linewidth]{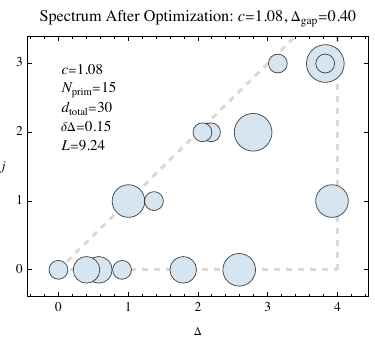}
    }$\qquad$
        \subfloat[][]{
    \includegraphics[width=0.4\linewidth]{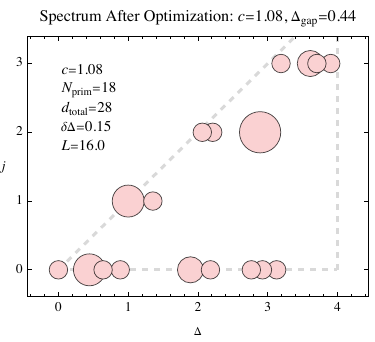}
    }\\
        \subfloat[][]{
    \includegraphics[width=0.4\linewidth]{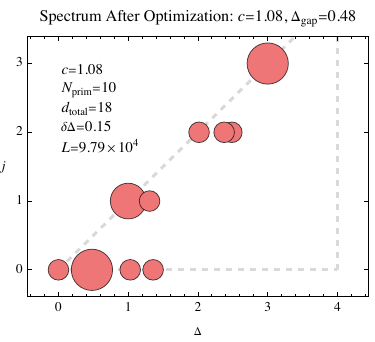}
    }$\qquad$
    \subfloat[][]{
    \includegraphics[width=0.4\linewidth]{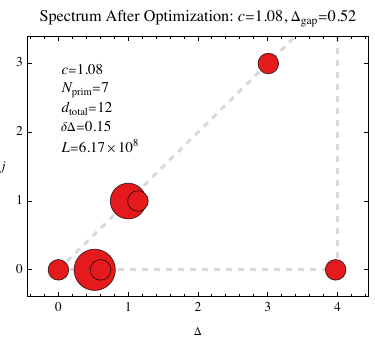}
    }
    \caption{Example primary spectra found in the $(c,\Delta_{\rm gap})$ survey, corresponding to CFT candidates with (a) high confidence, (b) medium confidence, and (c) low confidence, and poor optimization results with (d) weak exclusion, (e) intermediate exclusion, and (f) strong exclusion.
    These examples have $c = 1.08$ and $\Delta_{\rm gap} \in \{0.32,0.36,0.40, 0.44, 0.48, 0.52\}$, showing the transition from CFT-like to non-CFT-like behavior as the spectral gap increases for fixed central charge.
    }
    \label{fig:survey_examples}
\end{figure}

Finally, it is interesting to look at example spectra to see the transition from CFT-like to non-CFT-like behavior.
In \Fig{survey_examples}, we show 6 examples at $c = 1.08$, again combining operators with $|\Delta_a - \Delta_b| < 0.15$ for ease of visualization.
The CFT candidates with $\Delta_{\rm gap} \in (0.32,0.36,0.40)$ have the characteristic operator bunching behavior expected of CFTs.
The weakly-excluded spectrum with $\Delta_{\rm gap} = 0.44$ actually looks fairly reasonable, but its loss of $L = 16.045$ puts it \emph{just} above the $L = 16$ threshold we set for a CFT candidate.
The intermediately-excluded spectrum with $\Delta_{\rm gap} = 0.48$ is missing large twist operators expected of a true CFT.
The strongly-excluded spectrum is above the HCLY bound, and the lone high-twist operator at $(\Delta,j) = (4,0)$ is a remnant from a failed optimization.

\subsection{Descending Into the Gap}
\label{sec:descend_gap}

To better understand the origin and robustness of the apparent primal/dual gap, we perform two additional optimization studies aiming to identify what restrictions are responsible for generating this feature.

The first study is to relax the requirement of integer-valued degeneracy.
That is, we not only optimize the dimensions $\{\Delta_a\}$, but we also let the degeneracies $\{d_a\}$ float continuously between 0 and 2.
We focus on the gap region, and scan a grid of 45 $(c,\Delta_{\rm gap})$ pairs with:
\begin{align}
    c &\in \{1.02,1.04, 1.06, 1.08, 1.10\}, \nonumber \\ \Delta_{\rm gap} &\in \{0.24, 0.28, 0.32, 0.36, 0.40, 0.44, 0.48, 0.52, 0.56\}.
\label{eq:scan_region_gap}
\end{align}
Note that this is a computationally harder optimization problem, since there are now two parameters per primary operator instead of one, but because there are more pathways to a good minimum, we can still use the $\eta = 0.1$ learning rate with 300 iterations.

\begin{figure}
    \centering
\includegraphics[width=0.45\linewidth]{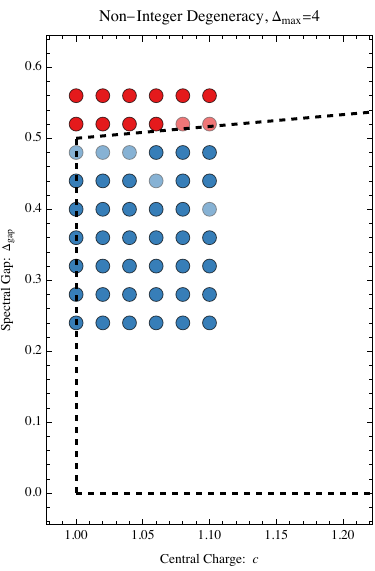}
    \caption{Scrutinizing the robustness of the $\Delta_{\rm max} = 4$ gap by doing a fresh optimization where operators are allowed to have non-integer degeneracy.
    Since there are explicit constructions that saturate the HCLY bound without integrality, we expect (and observe) the gap to close completely.
    }
    \label{fig:survey_non_integer_check}
\end{figure}

The result of the non-integral optimization is shown in \Fig{survey_non_integer_check}.
Without the integrality constraint, the good solutions fill the entire search space up to the HCLY bound.
As one gets closer to the bound, the quality of the optima tends to degrade, but still within the threshold of good solutions.
Since it is known how to saturate the HCLY bound with a non-integral spectrum~\cite{Fitzpatrick:2023lvh}, we expected the gap to close in this case.
Because we use the identical optimization code to handle the integer-restricted case, this is a strong indication that, if the primal/dual gap is real, it is due to imposing integrality.

\begin{figure}
    \centering  
  \subfloat[][]{
\includegraphics[width=0.45\linewidth]{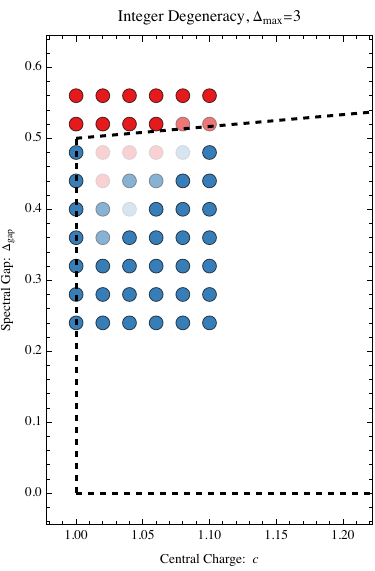}
\label{fig:survey_deltamax_3}
    }$\qquad$
    \subfloat[][]{
\includegraphics[width=0.45\linewidth]{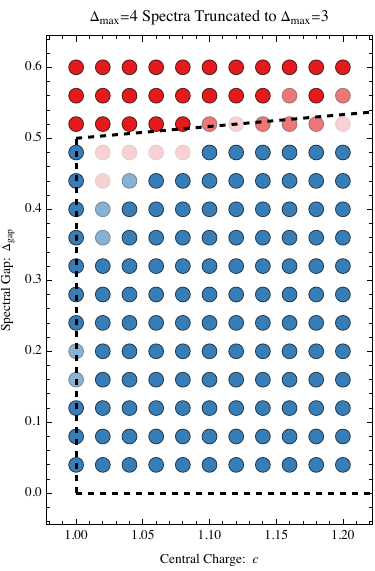}
\label{fig:survey_deltamax_4to3}
    }
    \caption{Scrutinizing the robustness of the $\Delta_{\rm max} = 4$ gap by relaxing the truncation scale to $\Delta_{\rm max} = 3$.
    In the left panel, we perform a fresh optimization and focus just on the gap region.
    In the right panel, we reinterpret the spectra already identified from the $\Delta_{\rm max} = 4$ survey from \Fig{survey_results}.
    In both cases, the gap closes significantly (though not completely), implying that the feature seen in the $\Delta_{\rm max} = 4$ case corresponds to constraints imposed by integrality from multiple $j = 2$ and $j = 3$ operators.
    The agreement between the two panels is strong evidence that the poor optimization results we found at $\Delta_{\rm max} = 4$ nevertheless have the right $\Delta_{\rm max} = 3$ structures.
    }
    \label{fig:survey_deltamax_3_checks}
\end{figure}

The second study is to see if the gap still appears in lower accuracy studies with a smaller $\Delta_{\rm max}$.
We focus on the same reduced scanning range from \Eq{scan_region_gap}, but with integer degeneracy restored.
We set $\Delta_{\rm max} = 3$ and reduce the number of primaries according to:
\begin{equation}
\label{eq:max3_num_operators}
    N_{j=0} = 18, \qquad     N_{j=1} = 12, \qquad     N_{j=2} = 6.
\end{equation}
We also reduce the cutoff in the fundamental domain to $\tau_2^{\rm max} = 5$, change the threshold for strong exclusion to $L = 10^5$, and use only $N_\tau^{\rm train} = 24$ training points and $k_{\rm SV} = 12$ singular values.
The reason we use fewer singular values is that with the default $k_{\rm SV} = 16$, the optimization was too aggressive about letting operators drift far above the $\Delta_{\rm max}$ threshold, causing it to sometimes end up in local minima.%
\footnote{An alternative strategy would be to decrease the learning rate $\eta$, but this is computationally more costly, whereas decreasing the number of singular values $k_{\rm SV}$ is computationally cheaper.}
This is a reminder of a key challenge of the numerical primal approach, which is that the failure to find a solution does not always indicate a fundamental obstruction to one existing.
In this particular case, though, we could easily adjust the optimization parameters to find better solutions.

The results of the $\Delta_{\rm max} = 3$ survey are shown in \Fig{survey_deltamax_3}.
We see that the gap closes significantly, though not entirely.
This suggests that the original gap is not just coming from a simple integrality constraint.
Instead, it seems that contributions from multiple spin $j \ge 2$ primaries are needed to create a large gap.
To validate this behavior, we take the spectra from the $\Delta_{\rm max} = 4$ survey, but analyze them with the relaxed $\Delta_{\rm max} = 3$ cut.
The results are shown in \Fig{survey_deltamax_4to3}, which agrees qualitatively with the results from the fresh optimization, suggesting that even when the optimizer finds poor results, the structures it does find correspond to lower $\Delta_{\rm max}$ behaviors.

The apparent importance of higher-spin operators for generating the gap is consistent with \Reference{Benjamin:2021ygh}, which conjectures that once the $j = 0$ and $j = 1$ spectra are fixed, the higher-spin spectra are completely determined.%
\footnote{When $c$ is greater than 25, it is also required to know the spectrum of light operators with $\Delta < \frac{c-1}{12}$, but that is not relevant for this study.}
The higher-spin primaries still need to have integer-valued degeneracies, though, which imposes constraints on the possible allowed $j = 0$ and $j = 1$ values.
When we set $\Delta_{\rm max} = 3$, there are only a few $j = 2$ operators, so it is apparently possible to still find good-enough solutions to the modular bootstrap equations.
With $\Delta_{\rm max} = 4$, we now have multiple $j = 2$ and $j = 3$ operators to consider, which is apparently enough to open up the gap.

\begin{figure}
    \centering
    \subfloat[]{
\includegraphics[width=0.45\linewidth]{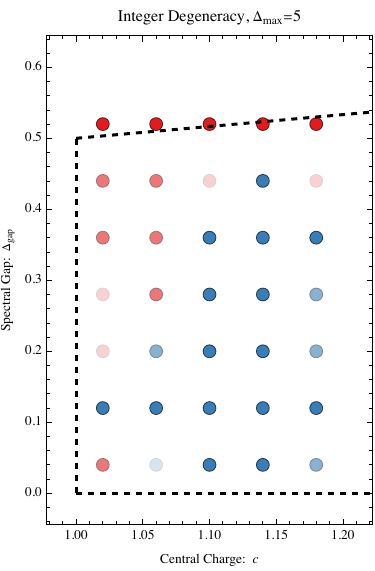}
    \label{fig:survey_deltamax_5}
    }$\qquad$
    \subfloat[]{
\includegraphics[width=0.45\linewidth]{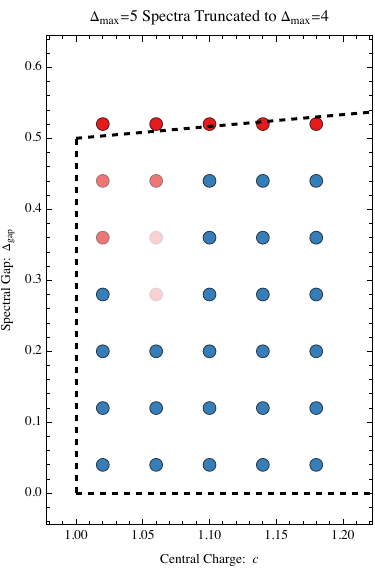}
    \label{fig:survey_deltamax_5to4}
    }
    \caption{Scrutinizing the robustness of the $\Delta_{\rm max} = 4$ gap by increasing the truncation threshold to $\Delta_{\rm max} = 5$.
    Because of computational limitations, we have to do a reduced scan in the $(c,\Delta_{\rm gap})$ plane.
    In the left panel, we show the results of a fresh optimization with $\Delta_{\rm max} = 5$.
    We find that the gap region persists in this $\Delta_{\rm max} = 5$ survey, though new poor optimization regions appear close to the HCLY bound and at $(c,\Delta_{\rm gap}) = (1.02,1.04)$.
    In the right panel, we reinterpret these spectra using the $\Delta_{\rm max} = 4$ criteria, finding the qualitatively same behavior as in the original $\Delta_{\rm max} = 4$ survey from \Fig{survey_results}.
    We therefore conclude that the apparent contraction of the space of candidate CFTs is due to new constraints appearing from integrality at $\Delta_{\rm max} = 5$.
    }
\end{figure}

\subsection{Impact of Additional Constraints}
\label{sec:deltamax5}

Based on the results of \Sec{descend_gap}, one might worry (or be excited, depending on your perspective) that if we were to raise the truncation scale, then the gap could even grow or change shape entirely.
In our final study, we set $\Delta_{\rm max} = 5$ and push the limits of our current optimization code.%
\footnote{While doing this study, we discovered a number of computational inefficiencies in our \mathematica\ code.  To ensure reproducibility of our results, we have not yet fixed these inefficiencies, but will do so for future studies.}
We increase the number of primary operators to:
\begin{equation}
\label{eq:max5_num_operators}
    N_{j=0} = 30, \qquad     N_{j=1} = 24, \qquad     N_{j=2} = 18, \qquad     N_{j=3} = 12, \qquad     N_{j=4} = 6,
\end{equation}
set the fundamental domain threshold to $\tau_2^{\rm max} = 7$, change the threshold for strong exclusion to $L = 10^{10}$, and increase the number of singular values and training samples to $k_{\rm SV} = 24$ and $N_\tau^{\rm train} = 48$, respectively.
To compensate for the increased computational cost, we roughly halve the number of $c$ and $\Delta_{\rm gap}$ values considered compared to the $\Delta_{\rm max} = 4$ case:
\begin{align}
    c &\in \{1.02, 1.06, 1.10, 1.14, 1.18\}, \nonumber \\ \Delta_{\rm gap} &\in \{0.04, 0.12, 0.20, 0.28, 0.36, 0.44, 0.52\}.
\label{eq:scan_region_deltamax5}
\end{align}
The results are shown in \Fig{survey_deltamax_5}.
Satisfyingly, the gap identified at $\Delta_{\rm max} = 4$ persists at $\Delta_{\rm max} = 5$.
There are, however, additional constraints on the spectrum at larger central charge, where some of the low confidence candidates at $\Delta_{\rm max} = 4$ are weakly excluded at $\Delta_{\rm max} = 5$.
There is also a poor optimization point at $(c,\Delta_{\rm gap}) = (1.02,1.04)$.
In \Fig{survey_deltamax_5to4}, we reinterpret these $\Delta_{\rm max} = 5$ spectra using the $\Delta_{\rm max} = 4$ criteria, and the results return qualitatively to the 
original $\Delta_{\rm max} = 4$ survey from \Fig{survey_results}.
This is strong evidence that additional constraints are appearing as we raise $\Delta_{\rm max}$, and we have likely not yet converged to the ultimate constraints possible from imposing integrality in the modular bootstrap.

The above cross checks explain why no hint of this primal/dual gap has previously appeared in the modular bootstrap literature.
Apart from \References{Kaidi:2020ecu,Fitzpatrick:2023lvh},  we are not aware of any studies of the impact of integrality in the context of generic 2d CFTs.
In \Reference{Fitzpatrick:2023lvh}, the impact of integrality was only studied for a few low-spin operators, and in \Reference{Kaidi:2020ecu}, precise statements were proved using integrality for currents in  rational CFTs and the space of cuspidal functions.
Since the primal/dual gap only opens up in going from $\Delta_{\rm max} = 3$ to $\Delta_{\rm max} = 4$, it appears that one must impose integrality on operators well above $\Delta_{\rm gap}$ to see the effect.
The fact that the space of candidate CFTs appears to further shrink at $\Delta_{\rm max} = 5$ indicates that there may be additional constraints from integrality of even higher-spin operators.
We hope this tantalizing result inspires new studies of spectral gap bounds in the dual modular bootstrap.

\section{Conclusions}
\label{sec:conclude}

In this paper, we used ML-style optimization to identify candidate 2d CFTs using a truncated and uncertainty-normalized version of the modular bootstrap equations.
After identifying five candidate spectra that satisfied our numerical definition of ``CFT-like,'' we used a parameter uncertainty analysis to claim evidence for a larger space of solutions to the modular bootstrap equations.
This suggests that, as far as modular invariance is concerned, it is straightforward to fill the central charge gap of $c \in (1,\frac{8}{7})$ between established CFTs.

We then performed a more comprehensive survey in the $(c, \Delta_{\rm gap})$ space of central charge versus spectral gap.
Intriguingly, we found strong evidence for a gap between our primal examples and the best known dual bound, with no examples found in the region $c \in (1.00,1.06)$ and $\Delta_{\rm gap} \in (0.3,0.5)$.
We argued that this primal/dual gap is driven by the constraint from unitarity that operator degeneracies are integer valued.
Moreover, this gap seems to depend on imposing integrality on operators with spin $j \ge 2$ and might even grow stronger with increasing $\Delta_{\rm max}$.
We hope these results inspire renewed efforts to study spectral gap bounds at $c \gtrsim 1$.

The two crucial ingredients in our numerical primal bootstrap approach were a principled way to estimate the impact of truncation and a powerful algorithm to navigate the loss landscape.
For truncation uncertainties, we used a novel deformation strategy to morph the partition function for a CFT with central charge $c=1$ into one that satisfies the Cardy scaling behavior of a CFT with $c > 1$.
For optimization algorithms, we used the recently developed \sven\ algorithm~\cite{Bright-Thonney:2026tlr} to navigate a complex, hierarchical loss landscape.
We suspect that our experience developing robust algorithms for uncertainty estimation and numerical optimization will be relevant for other theoretical physics problems involving hierarchical constraints.

The main open question from our study is whether the apparent primal/dual gap is actually real.
From the primal side, one would want to explore a broader range of optimization parameters and algorithms to gain more confidence in the difficulty of finding solutions in the gap.
Given the strong evidence for a continuous space of solutions, one would ideally directly navigate this  space to explore its boundaries, for example using generative modeling approaches.
From the dual side, one would want to extend the study of \Reference{Fitzpatrick:2023lvh} to impose integrality on higher-spin operators.
This would of course involve optimizing additional operators, which would grow the complexity of the problem.
In addition, if we want to probe possible constraints at small values of $\Delta_{\rm gap}$, we would likely need even more operators, since theories with smaller $\Delta_{\rm gap}$ are expected to have more relevant operators.

Coming back to the question from the introduction, does our study provide any insights into whether consistent small $c$ CFTs are rare or abundant?
From the perspective of the modular bootstrap, the evidence for a continuous family of solutions suggests that they are abundant.
To go beyond this statement, though, we will need to impose additional constraints on the properties of CFTs, in particular on crossing symmetry of the CFT correlation functions.
We anticipate that these crossing relations can also be translated into a ML language, enabling a primal approach to determining OPE coefficients from crossing symmetry relations.
Such an approach would allow us to descend even further into the space of  CFTs.

\acknowledgments{
JT thanks Samuel Bright-Thonney, Thomas Harvey, and Andre Lukas for collaborating on the development of \sven; Siddharth Mishra-Sharma for help arranging a Claude Max subscription;  James Halverson, Kevin Langhoff, Damien Leflot, Julian Sonner, Yotam Soreq, Andreas Stergiou, Ning Su, Piotr Tourkine, Yuan Xin, and Alexander Zhiboedov for helpful comments; and Nikolay Ebel, Rajeev Erramilli, Matthew Headrick, Dalimil Mazáč, Sridip Pal, Julio Parra-Martinez, Eric Perlmutter, Slava Rychkov, and other IHES cafeteria regulars for inspiring discussions.
JT is supported by the U.S.\ National Science Foundation (NSF) under Cooperative Agreement PHY-2019786 (The NSF AI Institute
for Artificial Intelligence and Fundamental Interactions,
\url{http://iaifi.org/}), by the U.S.\ Department of
Energy (DOE) Office of High Energy Physics under grant
number DE-SC0012567, by the Simons Foundation through Investigator grant 929241, and by the MIT Generative AI Impact Consortium (MGAIC).
JT also thanks the Institut des Hautes \'Etudes Scientifiques (IHES) and the Institut de Physique Th\'eorique (IPhT) for providing an inspiring sabbatical environment to carry out this research.
NB is supported by the U.S.\ Department of Energy (DOE) Office of High Energy Physics under grant number DE-SC0026324.
ALF and WL are supported by the
U.S.\ Department of Energy Office of Science under Award Number DE-SC0015845.
This work was initiated by NB and JT sharing an office at the Aspen Center for Physics, which is supported by National Science Foundation grant PHY-2210452.
Claude Opus 4.1 (and later 4.5 and 4.6) was used for background research for this project, but all the computations, text, and code for this paper were produced by the human authors, apart from final proofreading and the insight credited in footnotes~\ref{footnote:claude} and \ref{footnote:claude2}.
}

\appendix

\section{Alternative Deformation Strategy}
\label{app:alpha_truncation}

In this appendix, we explore an alternative deformation strategy that provides a complementary way to estimate truncation uncertainties.%
\footnote{In contrast to footnotes~\ref{footnote:claude} and \ref{footnote:claude2}, this approach was based entirely on human insights.}
It is conceptually simpler than the approach in \Sec{truncation}, but a bit more involved computationally, which is why we decided not to choose it as our default, though it may be worth using in future studies.

Like in \Sec{truncation}, the starting point is an exact CFT with central charge $c_0$ and partition function:
\begin{equation}
    Z_{\rm exact}(\tau;c_0).
\end{equation}
To build a deformed CFT with central charge $c$, we take $\alpha$ copies of the exact theory, with:
\begin{equation}
    \alpha = \frac{c}{c_0}.
\end{equation}
The resulting deformed theory is:
\begin{equation}
    Z_{\rm deformed}(\tau;c) = \Big(Z_{\rm exact}(\tau;c_0)\Big)^\alpha.
\end{equation}
If $\alpha$ were an integer, this would be a fully valid CFT partition function.
For non-integer $\alpha$, we still obtain a modular-invariant partition function, albeit one with fractional degeneracies.

For the baseline CFT in \Eq{baseline_cft} with $c_0 = 1$, its deformed partition function is simply:
\begin{equation}
Z^{\rm baseline}_{\rm deformed}(\tau; c) = \left(\frac{|\Theta_2(\tau)|^2 + |\Theta_3(\tau)|^2 + |\Theta_4(\tau)|^2}{2\, |\eta(\tau)|^2} \right)^c.
\end{equation}
It is instructive to expand this in the $q\to 0$ limit:
\begin{align}
\nonumber
    Z^{\rm baseline}_{\rm deformed}(\tau; c) &= |q|^{-\frac{c}{12}}\Big(1+ 2 \,c\, |q|^{\frac{1}{4}} + 2\,c\,(c-1) \,|q|^{\frac{1}{2}} + \frac{4}{3} \, c \, (c^2 - 3c + 2) \,|q|^{\frac{3}{4}} \\ &\qquad \qquad \qquad ~ + \frac{2}{3} \, c^2 \, (c^2 - 6c +11) \,|q| + c \, (q + \bar{q}) + \cdots
    \Big).
\end{align}
Compared to the analogous expression from \Eq{q_expanded_deformed}, this expression has many more terms, which can result in some computational headaches.
On the other hand, it does not have a logarithmic prefactor,
so this deformation yields a more physical spectrum in the sense of having a finite spectral gap.

To truncate this deformed partition function at $\Delta_{\rm max}$, we can write it as an expansion in $q$ and $\bar{q}$:
\begin{equation}
    Z_{\rm deformed}(\tau; c) = \frac{|q|^{\frac{1-c}{12}}}{|\eta(\tau)|^2}  \sum_{h, \bar{h}} \widetilde{a}_{h,\bar{h}} \, q^h \,  \bar{q}^{\bar{h}},
\end{equation}
where the coefficients $\widetilde{a}_{h,\bar{h}}$ will differ substantially from those of the exact CFT because of multiple cross terms induced by raising the partition function to the $\alpha$ power.
Unlike the expression in \Eq{deformed_via_coeffs}, there is no additional function of $\tau_2$, so it is straightforward to keep only the operators with $\Delta < \Delta_{\rm max}$:
\begin{equation}
    Z^{\rm trunc}_{\rm deformed}(\tau; c; \Delta_{\rm max}) = \frac{|q|^{\frac{1-c}{12}}}{|\eta(\tau)|^2}  \sum_{h, \bar{h}} \widetilde{a}_{h,\bar{h}} \, q^h \,  \bar{q}^{\bar{h}} \,\Theta\big(\Delta_{\rm max} - h - \bar{h}\big).
\end{equation}
Note that we did not have to perform any kind of inverse Laplace transform to obtain this expression and the spectrum is discrete.

\begin{table}[t]
\centering
\begin{tabular}{c | c  c c c c}
 & $m=5$ & $m=6$ & $m=7$ & $m=8$ & $m=9$ \\
\hline$R=1$ & $0.424$ & $0.449$ & $0.431$ & $0.428$ & $0.469$ \\
$\boldsymbol{R=\sqrt{2}}$ & $\mathbf{0.371}$ & $\mathbf{0.399}$ & $\mathbf{0.388}$ & $\mathbf{0.389}$ & $\mathbf{0.429}$ \\
$R=\sqrt{3}$ & $1.14$ & $1.04$ & $0.982$ & $0.957$ & $1.05$ \\
$R=2$ & $0.422$ & $0.453$ & $0.449$ & $0.455$ & $0.506$
\end{tabular}

\caption{
Loss values for the N1MM CFTs using an alternative strategy to estimate truncation uncertainties.
The loss values are around a factor of 2--3 smaller than for the default strategy in \Tab{n1mm_m_vs_r_table}, but otherwise quite similar in their qualitative patterns.
}
\label{tab:n1mm_m_vs_r_table_alpha}
\end{table}

To show the behavior of this alternative deformation strategy,
we recompute the loss values from \Tabs{n1mm_m_vs_r_table}{campaign_m_vs_c_table}.
We use the same uncertainty formula from \Eq{final_trunc_unc}, just with alternative definitions of $Z_{\rm deformed}$ and $Z^{\rm trunc}_{\rm deformed}$.
In \Tab{n1mm_m_vs_r_table_alpha}, we show the loss values for the N1MM models computed in this alternative way.
Apart from being roughly a factor of 2--3 smaller, the loss values exhibit the same qualitative features as the default approach.
In general, it seems that this alternative deformation strategy yields somewhat conservative uncertainties, and therefore  smaller loss values than the default.

\begin{table}
\centering
\begin{tabular}{c | c  c c c c}
 & $c=1.05$ & $c=1.07$ & $c=1.09$ & $c=1.11$ & $c=1.13$ \\
\hline$R=1$ & $0.870$ & $0.358$ & $0.416$ & $1.73$ & $1.31$ \\
$\boldsymbol{R=\sqrt{2}}$ & $\mathbf{0.725}$ & $\mathbf{0.305}$ & $\mathbf{0.360}$ & $\mathbf{1.50}$ & $\mathbf{1.15}$ \\
$R=\sqrt{3}$ & $3.42$ & $1.27$ & $1.43$ & $5.71$ & $4.46$ \\
$R=2$ & $0.866$ & $0.364$ & $0.421$ & $1.71$ & $1.31$
\end{tabular}

\caption{Loss values for the five candidate CFTs found in \Sec{hunt} using an alternative strategy to estimate truncation uncertainties.
Here, the loss values are around the same size as for the default strategy in \Tab{campaign_m_vs_c_table}, which makes sense since we did not specifically optimize using this alternative uncertainty strategy.
}
\label{tab:campaign_m_vs_c_table_alpha}
\end{table}

In \Tab{campaign_m_vs_c_table_alpha}, we show the alternative loss values for the five candidate CFTs identified in \Sec{hunt}.
Here, the loss values are quite similar to the default approach.
One has to keep in mind, though, that these candidates were identified using a loss function with the default uncertainties, not the alternative ones.
It is likely that if we reoptimized using the alternative uncertainties, the loss would decrease further.

\section{Additional Visualizations}
\label{app:additional_visualization}

In this appendix, we provide additional visualizations related to the hunt in \Sec{hunt}.

\begin{figure}
    \centering
    \subfloat[][]{
        \includegraphics[width=0.4\linewidth]{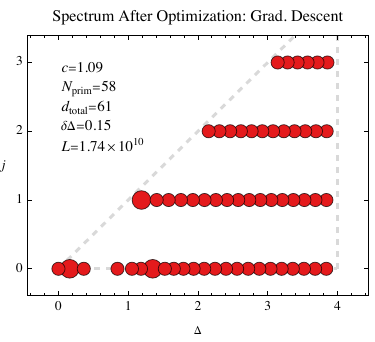}
    } $\qquad$
    \subfloat[][]{
        \includegraphics[width=0.4\linewidth]{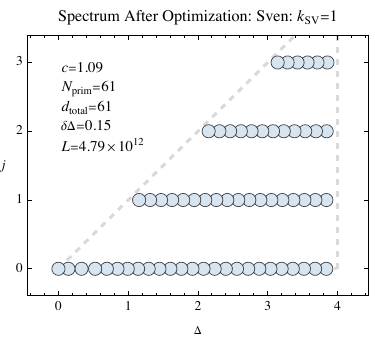}
    }\\
    \subfloat[][]{
        \includegraphics[width=0.4\linewidth]{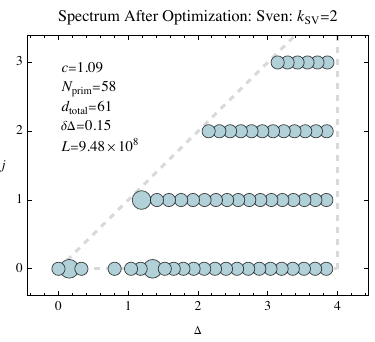}
    } $\qquad$
    \subfloat[][]{
        \includegraphics[width=0.4\linewidth]{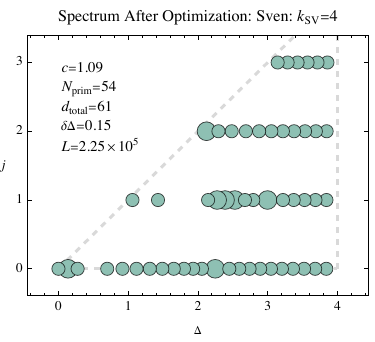}
    }\\
    \subfloat[][]{
        \includegraphics[width=0.4\linewidth]{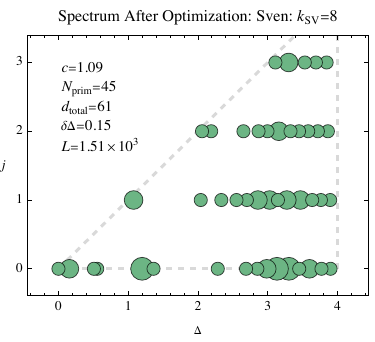}
    } $\qquad$
    \subfloat[][]{
        \includegraphics[width=0.4\linewidth]{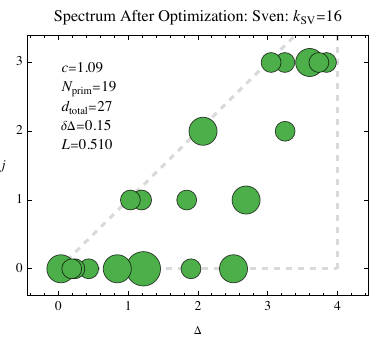}
        \label{fig:campaign_sv_scan_16}
    }
    \caption{Primary spectrum at the end of training, when using (a) gradient descent and (b--f) \sven\ with different numbers of singular values.
    The plots here correspond to the plateaus in \Fig{campaign_loss_vs_iter_svgrad}, and \Fig{campaign_sv_scan_16} is a duplicate of \Fig{campaign_ending_plot_109} for ease of comparison.
    To probe the spectrum up to $\Delta_{\rm max}$, it seems that roughly $2^{\Delta_{\rm max}}$ singular values are needed.
    }
    \label{fig:campaign_sv_scan}    
\end{figure}

In \Fig{campaign_sv_scan}, we show the different \sven\ optimization runs from \Fig{campaign_loss_vs_iter_svgrad} as we sweep over the number of singular values considered in the training.
Note that these training runs are independent, so one should be careful to avoid interpreting these as snapshots of the training dynamics.
Nevertheless, we see the hierarchical scales present in the problem.
Roughly speaking, when considering $k_{\rm SV}$ singular values, only operators with dimension below $\Delta_{\rm max} \approx \log_2 k_{\rm SV}$ are affected by the training before hitting a plateau in the loss landscape.

\begin{figure}[t]
    \centering
    \subfloat[][]{
        \includegraphics[width=0.45\linewidth]{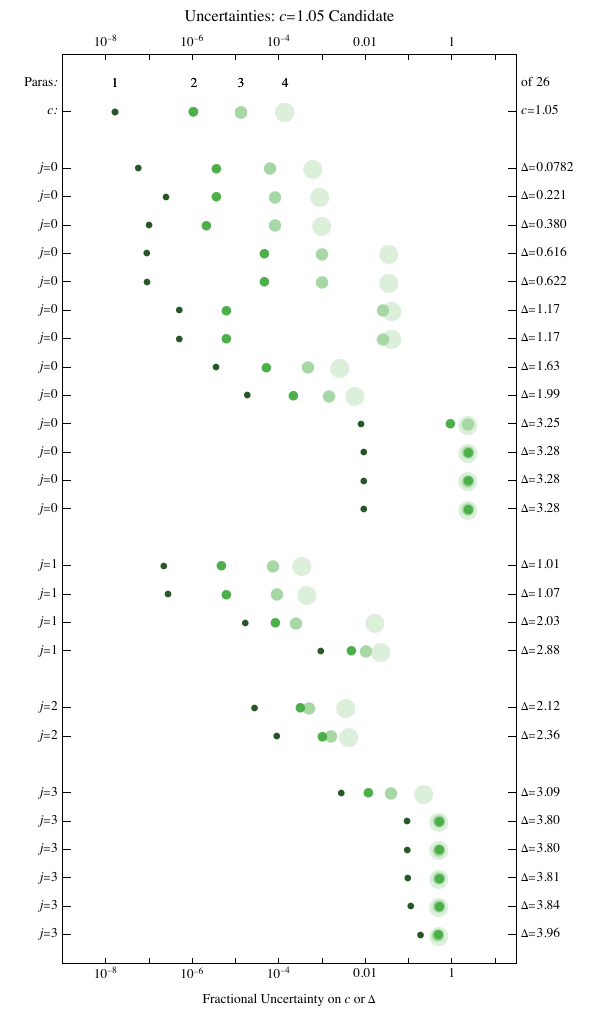}
    }$\qquad$
    \subfloat[][]{
        \includegraphics[width=0.45\linewidth]{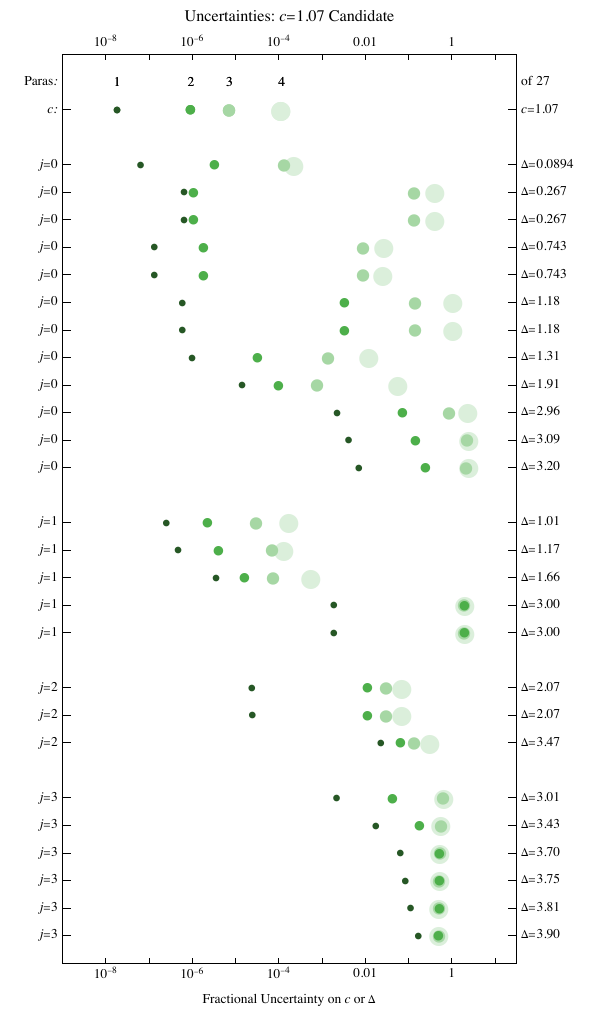}
    }
    \caption{Same as \Fig{campaign_uncertainty_109} but for the (a) $c = 1.05$ and (b) $c = 1.07$ candidates.}
    \label{fig:campaign_uncertainty_105_107}    
\end{figure}

\begin{figure}[t]
    \centering
    \subfloat[][]{
        \includegraphics[width=0.45\linewidth]{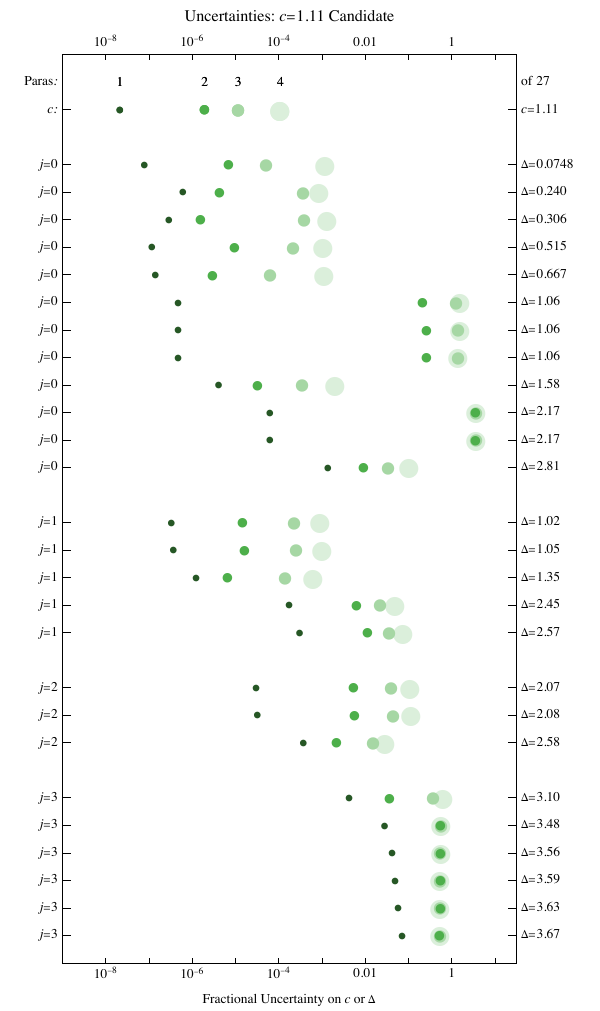}
    } $\qquad$
    \subfloat[][]{
        \includegraphics[width=0.45\linewidth]{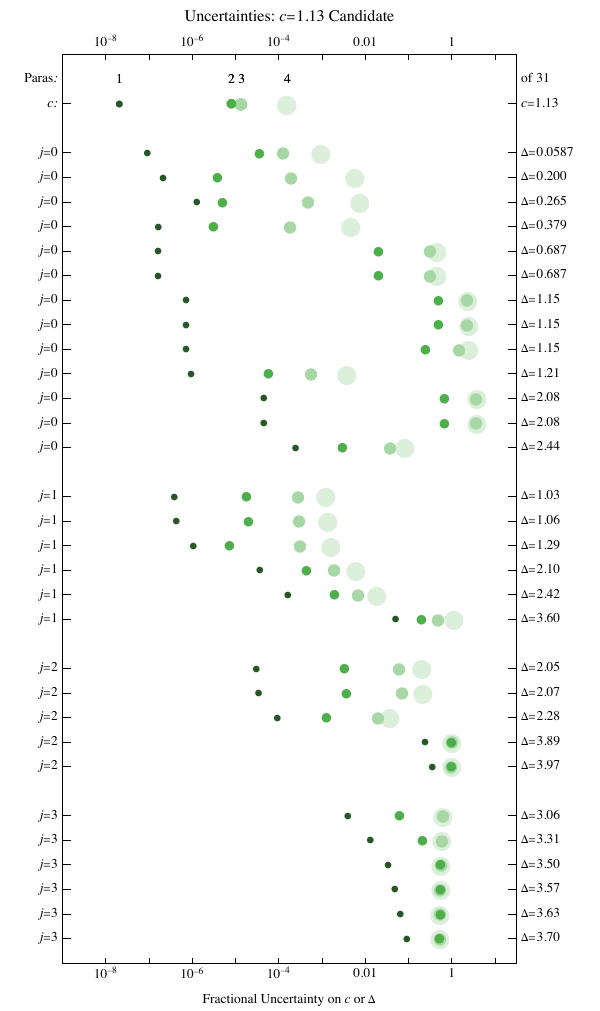}
    }
    
    \caption{Same as \Fig{campaign_uncertainty_109} but for the (a) $c = 1.11$ and (b) $c = 1.13$ candidates.}
    \label{fig:campaign_uncertainty_111_113}    
\end{figure}

In the main text, \Fig{campaign_uncertainty_109} showed the uncertainties on the inferred operator dimensions for the $c = 1.09$ candidate CFT.
For completeness, we show the uncertainties for the remaining four candidate models in \Figs{campaign_uncertainty_105_107}{campaign_uncertainty_111_113}, which exhibit the same qualitative features.
Specifically, the central charge and lower-dimension operators are more strongly constrained than the higher-dimension operators, but because of the strong correlations between parameters, the marginal uncertainties grow as multiple parameter variations are considered.

\FloatBarrier
\bibliographystyle{JHEP}
\bibliography{biblio.bib}

\end{document}